\documentclass[aps,prx,amsmath,superscriptaddress,twocolumn,bibnotes,longbibliography,floatfix]{revtex4-2}
\usepackage{amsmath}
\usepackage{amsfonts}
\usepackage{graphicx}
\usepackage{color}
\usepackage{dsfont}
\usepackage{verbatim}
\usepackage{mathtools}
\usepackage[normalem]{ulem}
\usepackage{comment}
\usepackage{stackrel}
\usepackage{bm}
\usepackage{booktabs}

\usepackage[colorlinks=true, 
linkcolor=blue, 
urlcolor=blue,
citecolor=blue]{hyperref}
\usepackage{orcidlink}
\usepackage[linesnumbered]{algorithm2e}

\RestyleAlgo{ruled}
\SetKwComment{Comment}{/* }{ */}
\DeclareMathOperator{\sgn}{sgn}

\definecolor{myblue}{rgb}{.93, .93, 1}

\newcommand{\bsub}{\begin{subequations}}
	\newcommand{\esub}{\end{subequations}}

\newcommand{\ket}[1]{\left| {#1} \right\rangle}
\newcommand{\bra}[1]{\left\langle {#1} \right|}

\begin{document}
	
	\title{Spin ladder quantum simulators from spin-orbit coupled quantum dot spin qubits}
	
	\author{Yang-Zhi~Chou~\orcidlink{0000-0001-7955-0918}}\email{yzchou@umd.edu}
	\affiliation{Condensed Matter Theory Center and Joint Quantum Institute, Department of Physics, University of Maryland, College Park, Maryland 20742, USA}

	\author{Sankar Das~Sarma}
	\affiliation{Condensed Matter Theory Center and Joint Quantum Institute, Department of Physics, University of Maryland, College Park, Maryland 20742, USA}
	
	\date{\today}

	\begin{abstract}
		Motivated by the recent Ge hole spin qubit experiments, we construct and study a two-leg spin ladder from a quantum dot array with spin-orbit couplings (SOCs), aiming to uncover the many-body phase diagrams and provide concrete guidance for the Ge hole spin qubit experiments. The spin ladder is described by an unprecedented, complex spin Hamiltonian, which contains antiferromagnetic Heisenberg exchange, Dzyaloshinskii–Moriya (DM), and anisotropic exchange interactions. We analyze the spin ladder Hamiltonian in two complementary situations, the strong rung coupling limit and the weak rung coupling limit. In the strong rung coupling limit, we systematically construct effective spin-$1/2$ chain models, connecting the well-studied one-dimensional spin models and providing a recipe for Hamiltonian engineering. It is worth emphasizing that effective DM interactions can be completely turned off while the microscopic DM interactions are generically inevitable. More intriguingly, the staggered DM interactions, which allow for the field-induced gap pertinent to the sine-Gordon model, can also be realized in the effective spin chain model, opening up a new route for studying the sine-Gordon model in spin qubits. In the weak rung coupling limit, we employ Abelian bosonization and Luther-Emery fermionization, uncovering a multitude of phases. Several commensurate-incommensurate transitions are driven by both the longitudinal magnetic field and the DM interactions in the legs (chains). Remarkably, the low-energy phase diagrams show strong dependence in the DM interaction, providing a concrete way to identify the strength of SOC in the experiments. We further demonstrate that different phases can be reached by tuning the magnetic field, both in amplitude and direction, paving the way for the efficient manipulation of the quantum many-body states in spin qubit experiments. Our work bridges quantum many-body theory and spin qubit device physics, establishing spin ladders made of spin-orbit-coupled quantum dots as a promising platform for engineering exotic spin models, constructing quantum many-body states, and paving the way for quantum computations and emulations.
	\end{abstract}
	
	\maketitle

	\section{Introduction}
	
	Spin qubits are a promising route for quantum technology, utilizing the spins of electrons or holes in semiconductor quantum dots or other systems at the nanometer scale \cite{LossD1998,HansonR2007,BurkardG2023}. A significant advantage is the compatibility with well-established semiconductor fabrication techniques, providing great potential for quantum computers with large-scale processors. Another advantage is that these quantum dot based spin qubits can be manipulated and operated entirely by electrical gate voltage control, making them both very fast and very precise. In addition, the gate-defined semiconductor quantum dot arrays, which are commonly used for spin qubits, are an ideal setup for exploring quantum many-body systems, such as Fermi Hubbard models \cite{HensgensT2017a,SinghaA2011,SalfiJ2016}, Nagaoka ferromagnetism \cite{DehollainJP2020,ButerakosD2019}, Heisenberg spin chains \cite{vanDiepenCJ2021}, spin ladders \cite{ZhangX2025,FarinaPC2025a}, and Coulomb drag \cite{HsiaoTK2024}. Thus, as was pointed out a long time ago, the quantum dot based spin qubit arrays are excellent analog quantum simulators for intractable solid state strongly correlated phenomena \cite{StaffordCA1994,KotlyarR1998,KotlyarR1998a}.
	
		\begin{figure}[t!]
		\includegraphics[width=0.4\textwidth]{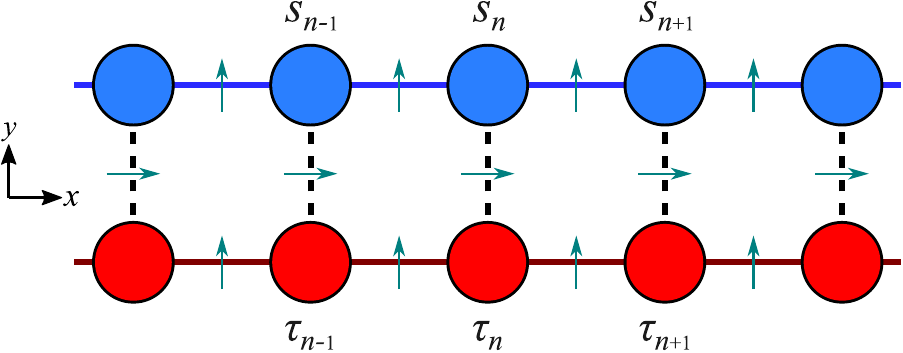}
		\caption{Setup of the two-leg spin ladder model. The blue (red) dots represent the $s$ ($\tau$) spins. The horizontal bonds indicate the interactions within a leg; the vertical bonds (black dashed lines) indicate the rung interactions. We use arrows in the bonds to indicate the direction of the DM vectors in the model. Notably, the DM vectors of the leg and rung couplings are orthogonal, realizing a complex spin model as we discuss in the main text.}
		\label{Fig:SL}
	\end{figure}

	Among the existing platforms for spin qubits, the Ge hole system \cite{WatzingerH2018,HendrickxNW2020,JirovecD2021,HendrickxNW2021a,JirovecD2022,BorsoiF2024,ZhangX2025,HsiaoTK2024,FarinaPC2025a,JirovecD2025} has attracted substantial recent interests because of its inherently strong spin-orbit coupling (SOC) \cite{vanRiggelen-DoelmanF2024,LiR2025}, nearly free from nuclear spin decoherence, low disorder and high mobility (55000 cm$^2/$V or even higher \cite{XieYH1993}), and the light effective mass. (See reviews \cite{ScappucciG2021,FangY2023} and references therein.) For example, $2\times 4$ \cite{ZhangX2025,HsiaoTK2024,FarinaPC2025a,JirovecD2025} and $4\times 4$ \cite{BorsoiF2024} Ge hole quantum dot arrays have been realized. The $2\times 4$ quantum dot array experiments are particularly interesting because of the ability to control and measure the dots individually. Moreover, the $2\times 4$ quantum dot arrays enable singlet-triplet encoding of spin qubits \cite{ZhangX2025,FarinaPC2025a,BurkardG2023} as well as exploring two-leg spin ladder physics \cite{GiamarchiT2003,GogolinAO2004}. 
	Due to the strong SOC in Ge hole systems, Dzyaloshinskii–Moriya (DM) interactions and anisotropic exchange interactions are no longer negligible \cite{MoriyaT1960b,MoriyaT1960d,ShekhtmanL1992,KavokinKV2001,KavokinKV2004,LiuZH2018}. However, the existing theoretical studies mainly focus on the isotropic antiferromagnetic (AF) Heisenberg exchange (arising from spin-preserving tunnelings) but rarely incorporate the DM and anisotropic exchange interactions, likely overlooking numerous nontrivial phenomena driven by strong SOCs. It is thus desirable to develop a systematic theory for the quantum dot arrays with significant SOCs, a largely unexplored topic. There are considerable current experimental efforts in using Ge hole based quantum dot spin qubits as solid state quantum simulator platforms where the interplay of SOC and interaction in strongly correlated systems can be directly studied in the laboratory for a nontrivial controlled Hamiltonian \cite{Demler}.

	In this work, we study a two-leg spin ladder made of quantum dots with SOCs as illustrated in Fig.~\ref{Fig:SL}, motivated by the recent Ge hole experiments \cite{ZhangX2025,HsiaoTK2024,FarinaPC2025a,JirovecD2025}. We focus on low-energy phases and Hamiltonian engineering in several analytically tractable limits, providing essential guidance for future experiments. First, we review and derive the spin-spin interactions based on a two-site Hubbard model with both spin-preserving and spin-flipping (Rashba-assisting) hoppings. In the single chain model, the DM interaction can be ``gauged away'' by local basis rotations \cite{DerzhkoO1994,ShekhtmanL1992,OshikawaM1997}, and the model in the rotated basis is described by a one-dimensional (1D) isotropic AF Heisenberg chain. In the two-leg spin ladder model, the DM vectors for the rung and the leg couplings are orthogonal to each other (as shown in Fig.~\ref{Fig:SL}), resulting in frustrations. Such a spin ladder is described by a complicated spin Hamiltonian, which, to the best of our knowledge, has not been discussed in the literature. We study this model in two complementary situations, strong and weak rung coupling limits. Both limits can be achieved experimentally by tuning the electric potentials, as demonstrated in the rung-based singlet-triplet encoding (strong rung coupling limit) \cite{ZhangX2025,FarinaPC2025a} and the Coulomb drag (nearly zero rung coupling limit) \cite{HsiaoTK2024} experimental setups.

	In the strong rung coupling limit, we solve the rung coupling exactly and derive effective spin-$1/2$ chains based on the low-lying states of the rung coupling Hamiltonian, following the ideas in Refs.~\cite{StrongSP1992,StrongSP1994,MilaF1998,ChaboussantG1998a,OrignacE2007}. This procedure reduces the Hilbert space dimension from $4^N$ to $2^N$ where $N$ is the number of rungs. The low-energy phases of the spin ladder can be inferred through the known 1D spin-$1/2$ model in the literature. A similar approach has been adopted in the recent studies on singlet-triplet encoding in Ge spin qubits \cite{ZhangX2025,FarinaPC2025a}. However, our calculations incorporate the DM and anisotropic interactions, offering a systematic theoretical framework for spin-orbit-coupled quantum dot spin ladders. Notably, the staggered DM interactions, which is key for realizing a nontrivial field-induced gap that can be described by the sine-Gordon model \cite{OshikawaM1997,AffleckI1999}, can be realized in the effective spin model through tuning the microscopic interactions. Our results indicate a novel way for studying various properties of the sine-Gordon model, making the spin-orbit-coupled quantum dot ladder an interesting analog quantum simulator. The key physics presented in the current work is the inclusion of the SOC explicitly in the Hamiltonian to predict the emergent quantum phase diagram appropriate for Ge hole quantum dot spin qubits.
	
	In the weak rung coupling limit, the rung couplings are treated perturbatively, and phase diagrams can be obtained analytically in a number of cases. Using Abelian bosonization and Luther-Emery fermionization \cite{GogolinAO2004,GiamarchiT2003}, we show that the magnetic field along the rung ($y$ direction in Fig.~\ref{Fig:SL}) and the angle of local basis rotation (for gauging away DM interactions in the legs) can induce several commensurate-incommensurate transitions \cite{PokrovskyVL1979}, resulting in a multitude of unprecedented phases that are absent in the regular spin ladder models with only AF Heisenberg interactions \cite{GogolinAO2004,GiamarchiT2003}. The phase diagrams depend on the DM interactions and thus provide a way to constrain the range of the unsettled SOC strength in the Ge hole spin qubit experiments. Aspects of our our predicted phase digram should be experimentally accessible in the currently available $2\times 4$ Ge hole spin qubit arrays.

	The two complementary approaches discussed above not only illustrate the low-energy quantum many-body phases but also provide hints for quantum simulations and quantum state preparation. The ability of engineering DM interactions makes the spin-orbit-coupled quantum dot spin ladder qualitatively distinct from other synthetic quantum simulators. Our results also demonstrate that a number of nontrivial phases can be realized by tuning the magnetic field, both in the magnitude and in the direction. These findings bridge the quantum many-body theory and the spin qubit device physics, establishing the spin-orbit-coupled quantum dot spin ladders as an interesting platform for studying exotic spin models, manipulating quantum many-body states, and enabling programmable quantum emulations.

	The rest of the paper is organized as follows: In Sec.~\ref{Sec:Model}, we review the microscopic spin-spin interactions and construct the spin models. In Sec.~\ref{Sec:strong_rung}, we analyze the two-leg spin ladder problem in the strong rung coupling limit, i.e., dominant $J_\perp$, $D_\perp$, and $\Gamma_\perp$. This approach provides a systematic way to reduce the problem to an effective spin-$1/2$ chain and allows for Hamiltonian engineering by tuning the microscopic spin-spin interactions. In Sec.~\ref{Sec:Weak_rung}, the rung couplings are treated as perturbations, and low-energy phase diagrams are constructed in several analytically tractable situations. The phase diagrams with a magnetic field along the $y$ direction are constructed using bosonization and Luther-Emery fermionization \cite{GogolinAO2004,GiamarchiT2003}; the phase diagrams with a magnetic field perpendicular to the $y$ direction are constructed based on the known single chain results \cite{GangadharaiahS2008,GarateI2010}. In Sec.~\ref{Sec:Discussion}, we discuss several technical remarks and experimental implications. In Appendix~\ref{App:Staggered_DM}, we provide an explicit derivation for realizing a staggered DM interaction in the effective spin chain. In Appendix~\ref{App:Bosonized_interaction}, the bosonized expressions for the rung couplings are derived. In Appendix \ref{App:LE}, we provide a derivation of the Luther-Emery fermionization, which is essential for determining the transition lines.

	\section{Model}\label{Sec:Model}
	
	We are interested in the spin models made of Ge hole-doped quantum dots. First, we review and discuss the microscopic interactions for a half-filled Hubbard model with SOC. Then, we focus on the corresponding the single spin chain and two-leg spin ladder models after projecting out the charge fluctuation in the Hubbard model. 
	
	\subsection{Inter-dot interaction}
	
	We consider quantum dot arrays in a 2D Ge hole spin qubit system \cite{WatzingerH2018,HendrickxNW2020,JirovecD2021,HendrickxNW2021a,JirovecD2022,BorsoiF2024,ZhangX2025,HsiaoTK2024,FarinaPC2025a,JirovecD2025}, which are known for their significant SOC. To describe the interactions between two quantum dots, we consider a single-orbital two-site spin-$\frac{1}{2}$ Hubbard model $\hat{H}_{\text{2-site}}=\hat{H}_{\text{dot}}+\hat{H}_{\text{hop}}+\hat{H}_{\text{SOC}}+\hat{H}_{\text{Zeeman}}$, where
	\begin{subequations}\label{Eq:H_2site}
		\begin{align}
			\nonumber\hat{H}_{\text{dot}}=&-\epsilon_1\sum_{s}c^{\dagger}_{1s}c_{1s}-\epsilon_2\sum_{s}c^{\dagger}_{2s}c_{2s}\\
			&+U\left(c^{\dagger}_{1\uparrow}c_{1\uparrow}c^{\dagger}_{1\downarrow}c_{1\downarrow}+c^{\dagger}_{2\uparrow}c_{2\uparrow}c^{\dagger}_{2\downarrow}c_{2\downarrow}\right),\\
			\hat{H}_{\text{hop}}=&-t\sum_{s}\left(c^{\dagger}_{1s}c_{2s}+\text{H.c.}\right)\\
			\hat{H}_{\text{SOC}}=&-r\sum_{s,s'}\left[ic^{\dagger}_{1s}\vec{\sigma}_{ss'}\cdot\left(\hat{r}_{12}\times\hat{z}\right)c_{2s'}+\text{H.c.}\right],\\
			\hat{H}_{\text{Zeeman}}=&-\vec{B}\cdot\overleftrightarrow{g}\cdot\sum_{s,s'}\left(c^{\dagger}_{1s}\vec{\sigma}_{ss'}c_{1s'}+c^{\dagger}_{2s}\vec{\sigma}_{ss'}c_{2s'}\right).
		\end{align}
	\end{subequations}
	In the above expressions, $-\epsilon_1$ and $-\epsilon_2$ are the single particle energies of the quantum dots, $s$ and $s'$ are the spin indexes, $c_{ns}$ is the electron annihilation operator of site $n$ with spin $s$, $U>0$ is the repulsive Hubbard interaction, $t>0$ encodes the strength of the spin-preserving hoppings, $r>0$ encodes the strength of the Rashba SOC-induced hoppings (e.g., Ref.~\cite{MutterPM2021}), $\hat{r}_{12}$ is the unit vector indicating the relative direction between quantum dots $1$ and $2$, $\vec{B}$ corresponds to the strength of magnetic field, and $\overleftrightarrow{g}$ is the g-factor matrix. In this work, we will assume that the g factor is a diagonal matrix, i.e., $\overleftrightarrow{g}=\text{diag}(g_x,g_y,g_z)$ for simplicity.
	
	In Eq.~(\ref{Eq:H_2site}), the $\hat{H}_{\text{dot}}$ describes the two isolated quantum dots. $\hat{H}_{\text{hop}}$ and $\hat{H}_{\text{SOC}}$ are the spin-preserving and Rashba SOC-induced hoppings \cite{MutterPM2021}, respectively. The $\hat{H}_B$ describes the Zeeman couplings between electrons and the external magnetic field. The parameters in Eq.~(\ref{Eq:H_2site}) can be obtained by ab initio calculations incorporating the quantum dot confining potentials and hole-doped Ge band structures. We are interested in the singly occupied states, corresponding to $0<\epsilon_1,\epsilon_2<U$. We focus on the limit $U\gg t,r,|\vec{B}|,\epsilon_1,\epsilon_2$ and treat $\hat{H}_{\text{hop}}$, $\hat{H}_{\text{SOC}}$, and $\hat{H}_B$ as perturbations. The derivation can be found in Refs.~\cite{MoriyaT1960b,MoriyaT1960d,ShekhtmanL1992,LiuZH2018}. We summarize the results in the following.
	
	The effective spin Hamiltonian between two spin qubits is given by
	\begin{align}
		\nonumber \hat{H}_{\text{2-spin}}=&J\vec{s}_1\cdot\vec{s}_2+D\hat{d}\cdot\left(\vec{s}_1\times\vec{s}_2\right)\\
		\nonumber&+\Gamma\left[2\left(\vec{s}_1\cdot\hat{d}\right)\left(\vec{s}_2\cdot\hat{d}\right)-\vec{s}_1\cdot\vec{s}_2\right]\\
		\label{Eq:H_2spin}&-\vec{B}\cdot\overleftrightarrow{g}\cdot\left(\vec{s}_1+\vec{s}_2\right),
	\end{align}
	where $\vec{s}_n$ denotes the spin$-\frac{1}{2}$ operator at site $n$, $J>0$ is the isotropic AF Heisenberg exchange interaction, $D>0$ indicates the strength of the DM interaction, and $\hat{d}=\hat{r}_{12}\times\hat{z}$ denotes the direction of the DM vector, and $\Gamma>0$ is the anisotropic interaction taking the form of the nearest neighbor dipole-dipole interaction. At the second-order perturbation theory, the coupling constants are given by \cite{MoriyaT1960b,MoriyaT1960d,ShekhtmanL1992,LiuZH2018,LiR2014}
	\begin{align}\label{Eq:couplings}
		J=\frac{4t^2}{U-\delta\epsilon^2/U},\,\,\,D=\frac{8tr}{U-\delta\epsilon^2/U},\,\,\,\Gamma=\frac{4r^2}{U-\delta\epsilon^2/U},
	\end{align}
	where $\delta\epsilon=\epsilon_1-\epsilon_2$. While the above spin-spin interaction expressions at the second order do not depend on the magnetic field, the magnetic field can contribute to spin-spin interactions at the order of $|\vec{B}|^3/U^3$ \cite{CookmeyerT2024}. Moreover, the tunneling matrix elements are known to depend on the magnetic field \cite{MutterPM2021}, resulting in dependence in the spin-spin interaction, also. For simplicity, we consider the second-order contribution based on the single-orbital Hubbard model, ignoring the explicit magnetic field dependence. The magnetic field corrections can be restored by modifying the values of the coupling constants.
	
	The DM interaction in Eq.~(\ref{Eq:H_2spin}) can be ``gauged away'' under a local basis rotation \cite{DerzhkoO1994,ShekhtmanL1992,OshikawaM1997}. To show this, we consider $\hat{d}=\hat{x}$ and introduce $\tilde{s}_1^x=s_1^x$, $\tilde{s}_2^y= s_x^y\cos\eta+ s_2^z\sin\eta$, and $\tilde{s}_2^z= s_x^z\cos\eta- s_2^y\sin\eta$, where $\eta=\tan^{-1}\left[D/(J-\Gamma)\right]$ is the angle of the rotation. The spin-spin interactions involving $y$ and $z$ components become
	\begin{align}
		\nonumber\hat{H}_{\text{2-spin}}\rightarrow&(J+\Gamma)s_1^x\tilde{s}_2^x+\sqrt{(J-\Gamma)^2+D^2}\left(s_1^y\tilde{s}_2^y+s_1^z\tilde{s}_2^z\right)\\
		\label{Eq:H_2spin_rotated}&-\vec{B}\cdot\overleftrightarrow{g}\cdot\left[\vec{s}_1+\mathcal{R}_x(\eta)\vec{\tilde{s}}_2\right],
	\end{align}
	where $\mathcal{R}_x(\eta)$ encodes the rotation of angle $\eta$ about the $x$ axis.
	The introduction of $\tilde{s}_2^{\mu}$ is equivalent to rotating the local basis of the site $2$ by an angle $-\eta$ about the $x$ axis. Note that the corresponding Zeeman coupling becomes rotated in the variable $\vec{\tilde{s}}_2$ due to the change of basis. The spin-spin interaction after the local basis rotation [the first line of Eq.~(\ref{Eq:H_2spin_rotated})] corresponds to an isotropic Heisenberg interaction because $\sqrt{(J-\Gamma)^2+D^2}=J+\Gamma$ based on the second-order degenerate perturbation theory results in Eq.~(\ref{Eq:couplings}). Such a local basis rotation is useful for certain lattice problems, such as the 1D chain (discussed next). However, for general situations such as the spin ladder model, the interaction bonds cannot be simplified simultaneously. The local basis rotations can still be useful in the limit that we treat certain bonds as perturbations as will be discussed later in this work.

	\subsection{Single spin chain}
	
	With the one-spin and two-spin interactions discussed previously, we can construct and study the many-body problem of spins. First, we consider a spin chain made of a linear array of Ge hole quantum dots. Specifically, we consider a translation-invariant chain along the $x$ direction. The spin chain Hamiltonian is given by
	\begin{align}
		\nonumber\hat{H}_{\text{chain}}=&\sum_n\left[\begin{array}{c}
			J\vec{s}_{n}\cdot\vec{s}_{n+1}+D\hat{y}\cdot\left(\vec{s}_{n}\times\vec{s}_{n+1}\right)\\[1mm]
			+\Gamma\left(2 s^y_{n}s^y_{n+1}-\vec{s}_{n}\cdot\vec{s}_{n+1}\right)
		\end{array}\right]\\
		\label{Eq:H_chain}&-\vec{B}\cdot\overleftrightarrow{g}\cdot\sum_n\vec{s}_n
	\end{align}
	where $n$ is the site index, $J$ denotes the AF Heisenberg interaction, $D$ denotes the DM interaction, and $\Gamma$ denotes the anisotropic (dipole-dipole) exchange interaction. We have used $\hat{d}=-\hat{x}\times\hat{z}=\hat{y}$, indicating the DM vector direction for each bond.
	
	For the spin chain given by Eq.~(\ref{Eq:H_chain}), one can gauge away DM interactions through local basis rotations about the $y$ axis \cite{DerzhkoO1994,ShekhtmanL1992,OshikawaM1997}, resulting in an isotropic AF Heisenberg interaction for each bond. The price we pay here is that the uniform magnetic field can become site-dependent. For a magnetic field along the $y$ direction, the magnetic field is unchanged after local basis rotations, and the problem corresponds to the well-known 1D AF Heisenberg model with a magnetic field \cite{MikeskaHJ2004}. For a magnetic field perpendicular to the $y$ direction, the problem has also been investigated in Ref.~\cite{GangadharaiahS2008,GarateI2010}. The main result is that a staggered magnetization along the $y$ direction can be realized along with a finite uniform magnetization along the field direction \cite{GarateI2010}. The results in the single chain model provide a foundation for weak rung coupling limit in Sec.~\ref{Sec:Weak_rung}.

	\subsection{Two-leg spin ladder}

	Now, we discuss a translation-invariant two-leg spin ladder made of Ge hole quantum dots, which is the main focus of this work. Specifically, we consider two coupled spin chains (along the $x$ direction) that are separated in the $y$ direction as illustrated in Fig.~\ref{Fig:SL}. The spin model is given by $\hat{H}_{\text{2-leg}}=\hat{H}_{\text{leg}}+\hat{H}_{\text{rung}}+\hat{H}_B$, where
	\begin{subequations}\label{Eq:H_spin_ladder}
		\begin{align}
			\nonumber\hat{H}_{\text{leg}}=&\sum_n\left[\begin{array}{c}
				J\vec{s}_{n}\cdot\vec{s}_{n+1}+D\hat{y}\cdot\left(\vec{s}_{n}\times\vec{s}_{n+1}\right)\\[1mm]
				+\Gamma\left(2 s^y_{n}s^y_{n+1}-\vec{s}_{n}\cdot\vec{s}_{n+1}\right)
			\end{array}\right]\\
			&+\sum_n\left[\begin{array}{c}
				J'\vec{\tau}_{n}\cdot\vec{\tau}_{n+1}+D'\hat{y}\cdot\left(\vec{\tau}_{n}\times\vec{\tau}_{n+1}\right)\\[1mm]
				+\Gamma'\left(2 \tau^y_{n}\tau^y_{n+1}-\vec{\tau}_{n}\cdot\vec{\tau}_{n+1}\right)
			\end{array}\right],\\
			\hat{H}_{\text{rung}}=&\sum_n\left[\begin{array}{c}
				J_{\perp}\vec{s}_{n}\cdot\vec{\tau}_{n}+D_{\perp}\hat{x}\cdot\left(\vec{s}_{n}\times\vec{\tau}_{n}\right)\\[1mm]
				+\Gamma_{\perp}\left(2 s^x_{n}\tau^x_{n}-\vec{s}_{n}\cdot\vec{\tau}_{n}\right)
			\end{array}\right],\\
			\hat{H}_B=&-\vec{B}\cdot\overleftrightarrow{g}\cdot\sum_n\left(\vec{s}_n+\vec{\tau}_n\right).
		\end{align}
	\end{subequations}
	In the above expressions, $\vec{s}_n$ ($\vec{\tau}_n$) represents the top (bottom) spin operator of the $n$th rung, $J$, $J'$, and $J_\perp$ denote the isotropic AF Heisenberg interactions, $D$, $D'$, and $D_\perp$ denote the DM interactions, and $\Gamma$, $\Gamma'$, and $\Gamma_\perp$ denote the anisotropic exchange interactions. As shown in Fig.~\ref{Fig:SL}, the DM vector directions are different for the horizontal bonds ($\hat{d}=\hat{y}$) and the vertical bonds ($\hat{d}=\hat{x}$) because of the Rashba SOC-induced hoppings. The primary magnetic field effect is the Zeeman coupling. An out-of-plane magnetic field (i.e., $z$ direction magnetic field) can induce fluxes and generate the plaquette interactions. However, these effects, higher-order contributions based on the Hubbard model, are likely small and thereby ignored in this work.

	The two-leg spin ladder model has several special limits. When the rung couplings are absent (i.e., $J_\perp=D_\perp=\Gamma_\perp=0$), the system is reduced to two decoupled spin chains. When the couplings in one of the legs are absent (e.g., $J'=D'=\Gamma'=0$), the two-leg spin ladder becomes a necklace model, related to the Kondo necklace model introduced by Doniach \cite{DoniachS1977}. Another special situation is a symmetric spin ladder with $J=J'$, $D=D'$, and $\Gamma=\Gamma'$, convenient for analytical approaches. 
	
	Unlike the spin chain, the local basis rotations cannot simplify all the interactions to isotropic AF Heisenberg interactions simultaneously, corresponding to the frustrated case as discussed in Ref.~\cite{ShekhtmanL1992}. Thus, the resulting two-leg spin ladder model is intrinsically complex under the generic situations. In this work, we focus on two complementary limits in which the effects of SOC can be simplified. In the strong rung coupling limit, we solve the two-spin problem for each vertical bond exactly and then treat the horizontal bonds perturbatively. In the converse situation, the weak rung coupling limit, we consider the decoupled chain limit and perform the local basis rotation, incorporating the SOC-induced terms along the legs. Then, we analyze the problem by treating the vertical bonds, the rung interactions, in a perturbative fashion.
	In the rest of the paper, we focus on the two-leg spin ladder model and discuss two different approaches in several analytically tractable cases.

	\section{Strong rung coupling limit: Effective spin-$1/2$ model}\label{Sec:strong_rung}

	In this section, we study the two-leg spin ladder in the strong rung coupling limit, i.e., $J_\perp,D_\perp,\Gamma_\perp\gg J,J',D,D',\Gamma,\Gamma'$. Following the ideas in Refs.~\cite{StrongSP1992,StrongSP1994,MilaF1998,ChaboussantG1998a,OrignacE2007}, we diagonalize $\hat{H}_{\text{rung}}+\hat{H}_B$ [given by Eqs.~(\ref{Eq:H_spin_ladder}b) and (\ref{Eq:H_spin_ladder}c)] exactly and focus on the two low-lying states for each rung. Then, we treat $\hat{H}_{\text{leg}}$ [given by Eq.~(\ref{Eq:H_spin_ladder}a)] as a perturbation and construct an effective spin-$1/2$ chain. The same approach has been adopted in the recent studies on singlet-triplet encoding in Ge spin qubits \cite{ZhangX2025,FarinaPC2025a} mainly based on the isotropic AF Heisenberg interactions. Here, our calculations incorporate the DM and anisotropic interactions, applicable to the spin-orbit-coupled quantum dot spin ladders generally.
	
	In the remainder of this section, we introduce the formalism for deriving the effective Hamiltonian. Then, we focus on several analytically tractable cases. The low-energy phases can be obtained by connecting the effective spin-$1/2$ model to the known results in the spin chain. Finally, we discuss how to manipulate the (staggered) DM interactions at the effective model level.

	\subsection{Formalism}
	
	Here, we discuss the formalism \cite{MilaF1998,ChaboussantG1998a,OrignacE2007} for constructing the effective spin-$1/2$ chain model in the strong rung coupling limit. First, we analyze the $\hat{H}_{\text{rung}}+\hat{H}_B$ [given by Eqs.~(\ref{Eq:H_spin_ladder}b) and (\ref{Eq:H_spin_ladder}c)], which describes decoupled rungs with a uniform external magnetic field. We can rewrite $\hat{H}_{\text{rung}}+\hat{H}_B=\sum_n\hat{H}_n$, where
	\begin{align}
		\nonumber\hat{H}_n=&J_{\perp}\vec{s}_{n}\cdot\vec{\tau}_{n}+D_{\perp}\hat{x}\cdot\left(\vec{s}_{n}\times\vec{\tau}_{n}\right)\\
		\label{Eq:H_n}&+\Gamma_{\perp}\left(2 s^x_{n}\tau^x_{n}-\vec{s}_{n}\cdot\vec{\tau}_{n}\right)-\vec{B}\cdot\overleftrightarrow{g}\cdot\left(\vec{s}_n+\vec{\tau}_n\right).
	\end{align}
	We assume $\overleftrightarrow{g}=\text{diag}(g_x,g_y,g_z)$ and consider $\vec{B}$ along certain high-symmetry directions. The eigenvectors and eigenvalues of $\hat{H}_n$ are given by $\psi_1$, $\psi_2$, $\psi_3$, $\psi_4$ and $E_1$, $E_2$, $E_3$, $E_4$, respectively. We are interested in the situations that $E_1,E_2\ll E_3,E_4$, where $E_1$ and $E_2$ are chosen to be the two lower energy states. Note that this situation can be satisfied for two special cases, (i) $D_\perp,\Gamma_\perp\ll J_\perp$ and (ii) $\vec{B}\parallel \hat{x}$, which we focus on in this work.
	
	When the lower energies $E_1$ and $E_2$ are well separated from the higher energies $E_3$ and $E_4$, one can construct a reduced low-energy Hilbert space spanned by $\psi_1$ and $\psi_2$. Formally, we require that $\Delta\equiv \min(E_3,E_4)-\max(E_1,E_2)$ is much larger than other perturbations, e.g., terms in $\hat{H}_{\text{leg}}$ [given by (\ref{Eq:H_spin_ladder}a)]. Within the reduced Hilbert space, we construct the local operators using the first-order perturbation theory as follows:
	\begin{align}\label{Eq:Local_op}
		\hat{O}\rightarrow\tilde{O}=&\left[\begin{array}{cc}
			\bra{\psi_1}\hat{O}\ket{\psi_1} & \bra{\psi_1}\hat{O}\ket{\psi_2}\\
			\bra{\psi_2}\hat{O}\ket{\psi_1} & \bra{\psi_2}\hat{O}\ket{\psi_2}
		\end{array}\right]_n.
	\end{align}
	For an operator $\hat{O}$, the corresponding operator in the reduced Hilbert space, $\tilde{O}$, can be expressed by a linear combination of Pauli matrices and the identity matrix.
	Note that $\tilde{C}$ (with $C=AB$) is generally not $\tilde{A}\tilde{B}$ because of the projection procedure. 
	The unperturbed local Hamiltonian $\hat{H}_n$ in the reduced Hilbert space is expressed by 
	\begin{align}
		\hat{H}_n\rightarrow\tilde{H}_n=&\left[\begin{array}{cc}
			\bra{\psi_1}\hat{H}_n\ket{\psi_1} & \bra{\psi_1}\hat{H}_n\ket{\psi_2}\\
			\bra{\psi_2}\hat{H}_n\ket{\psi_1} & \bra{\psi_2}\hat{H}_n\ket{\psi_2}
		\end{array}\right]_n\\
		\label{Eq:H_n_mapping}=&\left[\begin{array}{cc}
			E_1 & 0\\
			0 & E_2
		\end{array}\right]_n=-\frac{\delta E}{2}\sigma_n^z+\frac{E_{\text{tot}}}{2}I_n,
	\end{align}
	where $\delta E=E_2-E_1$, $E_{\text{tot}}=E_1+E_2$, $\sigma_n^z$ is the $z$ Pauli matrix, $I_n$ is the identity matrix, and the subscript $n$ denotes the site $n$. The ground state of $\tilde{H}_n$ is determined by $\delta E$, which plays the role of an external longitudinal magnetic field in Eq.~(\ref{Eq:H_n_mapping}). The $\psi_1$ ($\psi_2$) state corresponds to the spin-up (spin-down) state in the effective spin model. One can use Eq.~(\ref{Eq:Local_op}) to compute the corresponding operators for $\hat{H}_{\text{leg}}$ [given by Eq.~(\ref{Eq:H_spin_ladder}a)] and then construct an effective spin$-1/2$ model. The ground state of the effective spin chain is determined by the competition between the magnetic field coupling [Eq.~(\ref{Eq:H_n_mapping})] and other perturbations stemming from $\hat{H}_{\text{leg}}$. 
	
	Next, we construct the effective spin model in two analytically tractable limits: (i) Isotropic AF Heisenberg limit ($D_\perp,\Gamma_\perp\ll J_\perp$, treating $D_\perp,\Gamma_\perp$ as perturbation) and (ii) generic rung interactions with a $B_x$ field.

	\subsection{Isotropic AF Heisenberg interaction limit}\label{Sec:AF_rung}
	
	\begin{figure}[t!]
		\includegraphics[width=0.35\textwidth]{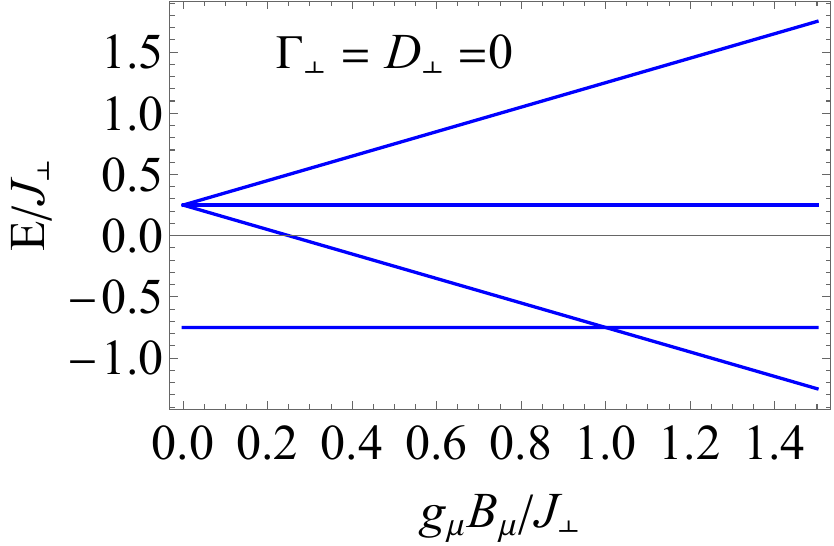}
		\caption{The energy spectrum of $\hat{H}_n$ as a function of magnetic field in the isotropic AF Heisenberg limit.}
		\label{Fig:AF_rung}
	\end{figure}
	
	For $J_\perp\gg D_\perp,\Gamma_\perp$, the main interactions in $\hat{H}_n$ are the isotropic AF Heisenberg interaction and the magnetic field coupling. In Fig.~\ref{Fig:AF_rung}, the energy spectrum as a function of magnetic field is plotted. In the absence of a magnetic field, the energy levels split into a ground state singlet and excited states triplets. The magnetic field brings down the energy of one of the triplets. Specifically, we choose $E_1=-3J_\perp/4$ and $E_2=J_\perp/4-g_\mu B_\mu$. The energy difference of the two low-lying states is given by $E_2-E_1=J_\perp-g_{\mu}B_{\mu}$, suggesting a transition at $g_{\mu}B_{\mu}=J_\perp$. For concreteness, The wavefunction $\psi_1$ is given $\frac{1}{\sqrt{2}}\left(\ket{\uparrow_s\downarrow_{\tau}}-\ket{\downarrow_s\uparrow_{\tau}}\right)$, and the wavefunction $\psi_2$ depends on the direction of the applied magnetic field. Note that our choice of basis $\psi_1$ and $\psi_2$ is different from Ref.~\cite{MilaF1998}, so the resulting local operator expressions differ by a unitary transformation $\tilde{O}'=\sigma^x_n\tilde{O}\sigma^x_n$. In addition, the DM ($D_\perp$) and anisotropic exchange ($\Gamma_\perp$) interactions in the rungs are treated perturbatively, providing further analytical understanding of the general situations. 
	
	In the rest of this subsection, we derive the effective spin-$1/2$ model for three particular cases: 1. $\vec{B}\parallel\hat{x}$, 2. $\vec{B}\parallel\hat{y}$, and 3. $\vec{B}\parallel\hat{z}$. The construction of the effective model here is valid as long as $g_\mu B_{\mu}\approx J_\perp$ and $\Delta\equiv \min(E_3,E_4)-\max(E_1,E_2)\gg J,J',D,D',D_\perp,\Gamma,\Gamma',\Gamma_\perp.$

	\subsubsection{$\vec{B}\parallel\hat{x}$}
	
	For the magnetic field along the $x$ direction (the DM vector direction of the rung couplings), the low-energy wavefunctions are given by $\psi_1=\frac{1}{\sqrt{2}}\left(\ket{\uparrow_s\downarrow_{\tau}}-\ket{\downarrow_s\uparrow_{\tau}}\right)$ and $\psi_2=\frac{1}{2}\left(\ket{\uparrow_s\uparrow_{\tau}}+\ket{\uparrow_s\downarrow_{\tau}}+\ket{\downarrow_s\uparrow_{\tau}}+\ket{\downarrow_s\downarrow_{\tau}}\right)$. With the low-energy states and Eq.~(\ref{Eq:Local_op}), we project the one-spin local operators onto the reduce Hilbert space as follows:
	\begin{subequations}
		\begin{align}
			s^x_n\rightarrow\frac{1}{4}\left(I_n-\sigma_n^z\right),\,\,\,s^y_n\rightarrow\frac{1}{2\sqrt{2}}\sigma_n^y,\,\,\,s^z_n\rightarrow\frac{1}{2\sqrt{2}}\sigma_n^x,\\
			\tau^x_n\rightarrow\frac{1}{4}\left(I_n-\sigma_n^z\right),\,\,\,\tau^y_n\rightarrow\frac{-1}{2\sqrt{2}}\sigma_n^y,\,\,\,\tau^z_n\rightarrow\frac{-1}{2\sqrt{2}}\sigma_n^x.
		\end{align}
	\end{subequations}
	Using the above expressions, we derive the contribution from $\hat{H}_{\text{leg}}$ [given by Eq.~(\ref{Eq:H_spin_ladder}a)]:
		\begin{align}
			\hat{H}_{\text{leg}}\rightarrow\sum_n\left\{\begin{array}{c}
				\!\!\frac{(J+J'-\Gamma-\Gamma')}{8}\left[\frac{(I_n-\sigma_n^z)(I_{n+1}-\sigma_{n+1}^z)}{2}+\sigma_n^x\sigma_{n+1}^x\right]\\[2mm]
				+\frac{(J+J'+\Gamma+\Gamma')}{8}\sigma_n^y\sigma_{n+1}^y\\[2mm]
				\!\!+\frac{D-D'}{8\sqrt{2}}\left[\sigma_n^x(I_{n+1}-\sigma^z_{n+1})-(I_n-\sigma^z_n)\sigma_{n+1}^x\right]
			\end{array}
			\right\}.
		\end{align}

	In addition, we project the $D_\perp$ and $\Gamma_\perp$ terms onto the reduced Hilbert space, equivalent to the first-order perturbation correction. The relevant two-spin local operators are mapped into
	\begin{subequations}
		\begin{align}
			&s^x_n\tau^x_n\rightarrow-\frac{1}{4}\sigma_n^z,\,\,\,\vec{s}_n\cdot\vec{\tau}_n\rightarrow-\frac{1}{2}\sigma_n^z-\frac{1}{4}I_n,\\
			&s^y_n\tau^z_n\rightarrow 0,\,\,\,s^z_n\tau^y_n\rightarrow 0.
		\end{align}
	\end{subequations}
	As a result, we obtain
	\begin{align}
		D_{\perp}\left(s_n^y\tau_n^z-s_n^z\tau_n^y\right)\rightarrow& 0,\\
		\Gamma_{\perp}\left(2 s^x_{n}\tau^x_{n}-\vec{s}_{n}\cdot\vec{\tau}_{n}\right)\rightarrow&\frac{\Gamma_\perp}{4}I_n.
	\end{align}
	The above results show trivial first-order contributions from $D_\perp$ and $\Gamma_\perp$ terms, indicating that $D_\perp$ and $\Gamma_\perp$ do not induce nontrivial effects in this limit. Note that the results here are distinct from other cases with a magnetic field perpendicular to $\hat{x}$. 
	
	With the results above, we can map the two-leg spin ladder into an effective spin-$1/2$ model (with $\mathcal{S}_n^{\mu}=\sigma_n^{\mu}/2$) given by
	\begin{align}\label{Eq:H_eff_AF_Bx}
		\nonumber\hat{H}_{\text{eff}}=&\sum_n\left[\begin{array}{c}
			\tilde{J}_x\mathcal{S}_n^x\mathcal{S}_{n+1}^x+\tilde{J}_y\mathcal{S}_n^y\mathcal{S}_{n+1}^y+\tilde{J}_z\mathcal{S}_n^z\mathcal{S}_{n+1}^z\\[1mm]
			+\tilde{D}_y\left(\mathcal{S}_n^z\mathcal{S}_{n+1}^x-\mathcal{S}_n^x\mathcal{S}_{n+1}^z\right)-\tilde{h}_z\mathcal{S}^z_n
		\end{array}
		\right]\\
		&+\frac{\tilde{J}_z}{2}\left(\mathcal{S}_{n_L}+\mathcal{S}_{n_R}\right)+\frac{\tilde{D}_y}{2}\left(\mathcal{S}^x_{n_L}-\mathcal{S}^x_{n_R}\right),
	\end{align}
	where $n_R$ and $n_L$ are the right and left boundary sites,
	\begin{subequations}\label{Eq:couplings_Bx_AF}
		\begin{align}
			\tilde{J}_x=&(J+J'-\Gamma-\Gamma')/2,\\
			\tilde{J}_y=&(J+J'+\Gamma+\Gamma')/2,\\
			\tilde{J}_z=&(J+J'-\Gamma-\Gamma')/4,\\
			\tilde{D}_y=&(D-D')/2\sqrt{2},\\
			\tilde{h}_z=&J_\perp-g_xB_x+(J+J'-\Gamma-\Gamma')/4.
		\end{align}
	\end{subequations}
	The second line of Eq.~(\ref{Eq:H_eff_AF_Bx}) describes the boundary terms with $n_R$ ($n_L$) being the rightmost (leftmost) site.
	The resulting effective Hamiltonian describes a 1D XYZ model with a DM interaction and an external magnetic field along the $z$ direction.  Notably, $\tilde{D}_y$ can be completely turned off by setting $D=D'$.
	
	It is worth emphasizing that the $\tilde{h}_z$ term acquires contributions from the leg coupling [i.e., the $(J+J'-\Gamma-\Gamma')/4$ part in Eq.~(\ref{Eq:couplings_Bx_AF})]. The origin of such a correction originates from the projected one-spin local operator, $s^x_n,\tau_n^x\rightarrow\left(I_n-\sigma_n^z\right)/4$, which projects out the up-spin state entirely. As a consequence, several two-(pseudo)spin terms have zero contributions as long as one of the $\sigma$-spin is up. The contributions to $\tilde{h}_z$ reflect the nature of the projection operator. Similar situations can also be found in other cases.
	
	\subsubsection{$\vec{B}\parallel\hat{y}$}

For the magnetic field along the $y$ direction (the DM vector direction of the leg couplings), the low-energy wavefunctions are given by $\psi_1=\frac{1}{\sqrt{2}}\left(\ket{\uparrow_s\downarrow_{\tau}}-\ket{\downarrow_s\uparrow_{\tau}}\right)$ and $\psi_2=\frac{1}{2}\left(-\ket{\uparrow_s\uparrow_{\tau}}-i\ket{\uparrow_s\downarrow_{\tau}}-i\ket{\downarrow_s\uparrow_{\tau}}+\ket{\downarrow_s\downarrow_{\tau}}\right)$. With the low-energy states and Eq.~(\ref{Eq:Local_op}), the projected one-spin local operators are given by
	\begin{subequations}
		\begin{align}
			s^x_n\rightarrow\frac{1}{2\sqrt{2}}\sigma_n^x,\,\,\,s^y_n\rightarrow\frac{1}{4}\left(I_n-\sigma_n^z\right),\,\,\,s^z_n\rightarrow\frac{1}{2\sqrt{2}}\sigma_n^y,\\
			\tau^x_n\rightarrow\frac{-1}{2\sqrt{2}}\sigma_n^x,\,\,\,\tau^y_n\rightarrow\frac{1}{4}\left(I_n-\sigma_n^z\right),\,\,\,\tau^z_n\rightarrow\frac{-1}{2\sqrt{2}}\sigma_n^y,
		\end{align}
	\end{subequations}
	Next, we derive the contribution from $\hat{H}_{\text{leg}}$ [given by Eq.~(\ref{Eq:H_spin_ladder}a)]  as follows:
	\begin{align}
		\hat{H}_{\text{leg}}\rightarrow\sum_n\left\{\begin{array}{c}
			\frac{(J+J'-\Gamma-\Gamma')}{8}\left(\sigma_n^x\sigma_{n+1}^x+\sigma_n^y\sigma_{n+1}^y\right)\\[1mm]
			\!\!+\frac{(J+J'+\Gamma+\Gamma')}{16}(I_n-\sigma_n^z)(I_{n+1}-\sigma_{n+1}^z)\\[1mm]
			-\frac{D+D'}{8}\left[\sigma_n^x\sigma_{n+1}^y-\sigma_n^y\sigma_{n+1}^x\right]
		\end{array}
		\right\}.
	\end{align}
	
	We can also treat $D_\perp$ and $\Gamma_\perp$ terms in the first-order perturbation theory. The relevant two-spin local operators become
	\begin{subequations}
		\begin{align}
			&s^x_n\tau^x_n\rightarrow-\frac{1}{8}(I_n+\sigma_n^z),\,\,\,\vec{s}_n\cdot\vec{\tau}_n\rightarrow-\frac{1}{2}\sigma_n^z-\frac{1}{4}I_n,\\
			&s^y_n\tau^z_n\rightarrow \frac{-1}{4\sqrt{2}}\sigma_n^y,\,\,\,s^z_n\tau^y_n\rightarrow \frac{1}{4\sqrt{2}}\sigma_n^y.
		\end{align}
	\end{subequations}
	As a result, we obtain
	\begin{align}
		D_{\perp}\left(s_n^y\tau_n^z-s_n^z\tau_n^y\right)\rightarrow& -\frac{D_\perp}{2\sqrt{2}}\sigma_n^y,\\
		\Gamma_{\perp}\left(2 s^x_{n}\tau^x_{n}-\vec{s}_{n}\cdot\vec{\tau}_{n}\right)\rightarrow&\frac{\Gamma_\perp}{4}\sigma_n^z.
	\end{align}
	The role of $D_\perp$ is to induce hybridization between $\psi_1$ and $\psi_2$, avoiding level crossing; the $\Gamma_\perp$ acts as an effective magnetic field along the $-z$ direction.
	
	With the results above, we can map the two-leg spin ladder into an effective spin-$1/2$ model (with $\mathcal{S}_n^{\mu}=\sigma_n^{\mu}/2$) given by
	\begin{align}\label{Eq:H_eff_AF_By}
		\nonumber\hat{H}_{\text{eff}}=&\sum_n\left[\begin{array}{c}
			\tilde{J}_{xy}\left(\mathcal{S}_n^x\mathcal{S}_{n+1}^x+\mathcal{S}_n^y\mathcal{S}_{n+1}^y\right)+\tilde{J}_z\mathcal{S}_n^z\mathcal{S}_{n+1}^z\\[1mm]
			+\tilde{D}_z\left(\mathcal{S}_n^x\mathcal{S}_{n+1}^y-\mathcal{S}_n^y\mathcal{S}_{n+1}^x\right)-\tilde{h}_y\mathcal{S}^y_n-\tilde{h}_z\mathcal{S}^z_n
		\end{array}
		\right]\\
		&+\frac{\tilde{J}_z}{2}\left(\mathcal{S}^z_{n_L}+\mathcal{S}^z_{n_R}\right),
	\end{align}
	where $n_R$ and $n_L$ are the right and left boundary sites,
	\begin{subequations}\label{Eq:couplings_By_AF}
		\begin{align}
			\tilde{J}_{xy}=&(J+J'-\Gamma-\Gamma')/2,\\
			\tilde{J}_z=&(J+J'+\Gamma+\Gamma')/4,\\
			\tilde{D}_z=&-(D+D')/2,\\
			\tilde{h}_y=&D_\perp/\sqrt{2},\\
			\tilde{h}_z=&J_\perp-g_yB_y+(J+J'+\Gamma+\Gamma')/4-\Gamma_\perp/2.
		\end{align}
	\end{subequations}
	The second line of Eq.~(\ref{Eq:H_eff_AF_By}) describes the boundary terms.
	The effective spin-$1/2$ model is an XXZ spin chain with a DM interaction and external magnetic fields along both the $y$ and $z$ directions, which has been studied in Refs.~\cite{GangadharaiahS2008,GarateI2010}. Unlike other cases, the DM interaction $\tilde{D}_z$ cannot be turned off in case as $D,D'> 0$ generically.

	\subsubsection{$\vec{B}\parallel\hat{z}$}
	
	For the magnetic field along the $z$ direction (the out-of-plane direction), the low-energy wavefunctions are given by $\psi_1=\frac{1}{\sqrt{2}}\left(\ket{\uparrow_s\downarrow_{\tau}}-\ket{\downarrow_s\uparrow_{\tau}}\right)$ and $\psi_2=\ket{\uparrow_s\uparrow_{\tau}}$. With the low-energy states and Eq.~(\ref{Eq:Local_op}), the projected one-spin local operators are expressed by
	\begin{subequations}
		\begin{align}
			s^x_n\rightarrow\frac{-1}{2\sqrt{2}}\sigma_n^x,\,\,\,s^y_n\rightarrow\frac{1}{2\sqrt{2}}\sigma_n^y,\,\,\,s^z_n\rightarrow\frac{1}{4}\left(I_n-\sigma_n^z\right),\\
			\tau^x_n\rightarrow\frac{1}{2\sqrt{2}}\sigma_n^x,\,\,\,\tau^y_n\rightarrow\frac{-1}{2\sqrt{2}}\sigma_n^y,\,\,\,\tau^z_n\rightarrow\frac{1}{4}\left(I_n-\sigma_n^z\right).
		\end{align}
	\end{subequations}
	Note that our choice of basis $\psi_1$ and $\psi_2$ is different from Ref.~\cite{MilaF1998}, so the resulting local operator expressions differ by a unitary transformation $\tilde{O}'=\sigma^x_n\tilde{O}\sigma^x_n$.
	Next, we derive the contribution from $\hat{H}_{\text{leg}}$ [given by Eq.~(\ref{Eq:H_spin_ladder}a)]  as follows:
	\begin{align}
		\hat{H}_{\text{leg}}\rightarrow\sum_n\left\{\begin{array}{c}
			\!\!\frac{(J+J'-\Gamma-\Gamma')}{8}\!\left[\sigma_n^x\sigma_{n+1}^x+\frac{(I_n-\sigma_n^z)(I_{n+1}-\sigma_{n+1}^z)}{2}\right]\!\!\\[2mm]
			+\frac{(J+J'+\Gamma+\Gamma')}{8}\sigma_n^y\sigma_{n+1}^y\\[2mm]
			\!\!-\frac{D-D'}{8\sqrt{2}}\!\left[(I_n-\sigma^z_n)\sigma_{n+1}^x-\sigma_n^x(I_{n+1}-\sigma^z_{n+1})\right]\!\!
		\end{array}
		\right\}.
	\end{align}
	We can also treat $D_\perp$ and $\Gamma_\perp$ in the first-order perturbation theory. The relevant two-spin local operators become
	\begin{subequations}
		\begin{align}
			&s^x_n\tau^x_n\rightarrow-\frac{1}{8}(I_n+\sigma_n^z),\,\,\,\vec{s}_n\cdot\vec{\tau}_n\rightarrow-\frac{1}{2}\sigma_n^z-\frac{1}{4}I_n,\\
			&s^y_n\tau^z_n\rightarrow \frac{1}{4\sqrt{2}}\sigma_n^y,\,\,\,s^z_n\tau^y_n\rightarrow \frac{-1}{4\sqrt{2}}\sigma_n^y.
		\end{align}
	\end{subequations}
	As a result, we obtain
	\begin{align}
		D_{\perp}\left(s_n^y\tau_n^z-s_n^z\tau_n^y\right)\rightarrow& \frac{D_\perp}{2\sqrt{2}}\sigma_n^y,\\
		\Gamma_{\perp}\left(2 s^x_{n}\tau^x_{n}-\vec{s}_{n}\cdot\vec{\tau}_{n}\right)\rightarrow&\frac{\Gamma_\perp}{4}\sigma_n^z.
	\end{align}
	The role of $D_\perp$ is to hybridize $\psi_1$ and $\psi_2$, avoiding level crossing; the $\Gamma_\perp$ acts as an effective magnetic field along the $-z$ direction. The effects here are similar to the $\vec{B}\parallel \hat{y}$ case.
	
	With the results above, we can map the two-leg spin ladder into an effective spin-$1/2$ model (with $\mathcal{S}_n^{\mu}=\sigma_n^{\mu}/2$) given by
	\begin{align}\label{Eq:H_eff_AF_Bz}
		\nonumber\hat{H}_{\text{eff}}=&\sum_n\left[\begin{array}{c}
			\tilde{J}_x\mathcal{S}_n^x\mathcal{S}_{n+1}^x+\tilde{J}_y\mathcal{S}_n^y\mathcal{S}_{n+1}^y+\tilde{J}_z\mathcal{S}_n^z\mathcal{S}_{n+1}^z\\[1mm]
			+\tilde{D}_y\left(\mathcal{S}_n^z\mathcal{S}_{n+1}^x-\mathcal{S}_n^x\mathcal{S}_{n+1}^z\right)-\tilde{h}_y\mathcal{S}^y_n-\tilde{h}_z\mathcal{S}^z_n
		\end{array}
		\right]\\
		&+\frac{\tilde{J}_z}{2}\left(\mathcal{S}^z_{n_L}+\mathcal{S}^z_{n_R}\right)+\frac{\tilde{D}_y}{2}\left(\mathcal{S}^x_{n_L}-\mathcal{S}^x_{n_R}\right),
	\end{align}
	where $n_R$ and $n_L$ are the right and left boundary sites,
	\begin{subequations}\label{Eq:couplings_Bz_AF}
		\begin{align}
			\tilde{J}_x=&(J+J'-\Gamma-\Gamma')/2,\\
			\tilde{J}_y=&(J+J'+\Gamma+\Gamma')/2,\\
			\tilde{J}_z=&(J+J'-\Gamma-\Gamma')/4,\\
			\tilde{D}_y=&(D-D')/2\sqrt{2},\\
			\tilde{h}_y=&-D_\perp/\sqrt{2},\\
			\tilde{h}_z=&J_\perp-g_zB_z+(J+J'-\Gamma-\Gamma')/4-\Gamma_\perp.
		\end{align}
	\end{subequations}
	The second line of Eq.~(\ref{Eq:H_eff_AF_By}) describes the boundary terms.
	The effective spin-$1/2$ model is an XYZ spin chain with a DM interaction and external magnetic fields along both the $y$ and $z$ directions. We note that $\tilde{D}_y$ can be completely turned off by setting $D=D'$.
	
	\subsection{General rung coupling}
	
	\begin{figure}[t!]
		\includegraphics[width=0.35\textwidth]{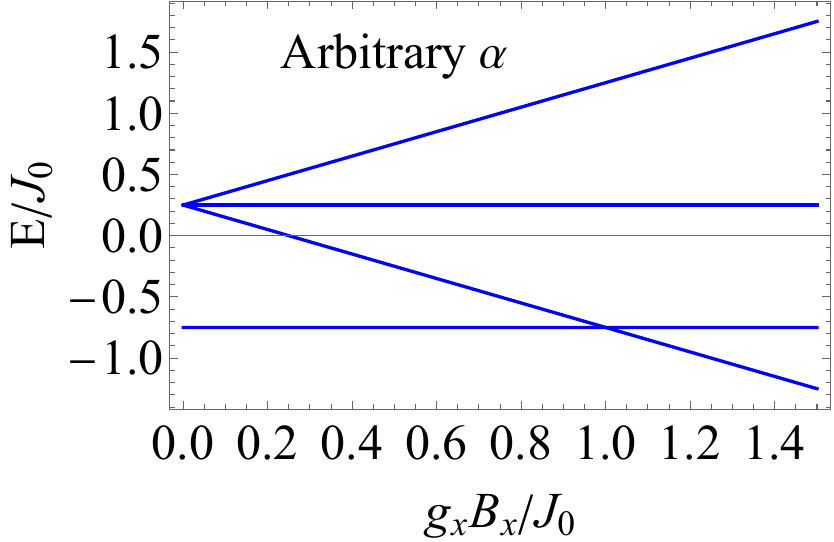}
		\caption{The energy spectrum of $\hat{H}_n$ [Eq.~(\ref{Eq:H_gen_rung})] under a magnetic field along $\hat{x}$. The results here are identical to Fig.~\ref{Fig:AF_rung}.}
		\label{Fig:General_rung_Bx}
	\end{figure}

	For the general situations, we can parametrize the coupling constants as follows: $J_{\perp}=J_0\cos^2\alpha$, $\Gamma_{\perp}=J_0\sin^2\alpha$, and $D_{\perp}=2J_0\sin\alpha\cos\alpha$, where $0\le\alpha\le\pi/2$. The parametrization here is generally valid for the quantum dots with SOCs as discussed in Sec.~\ref{Sec:Model}. The local Hamiltonian reads
	\begin{align}
		\label{Eq:H_gen_rung}\hat{H}_n=J_0\left[\begin{array}{c}
			\cos^2\alpha\,\vec{s}_{n}\cdot\vec{\tau}_{n}+2\cos\alpha\sin\alpha\,\hat{x}\cdot\left(\vec{s}_{n}\times\vec{\tau}_{n}\right)\\[1mm]
			+\sin^2\alpha\,\left(2 s^x_{n}\tau^x_{n}-\vec{s}_{n}\cdot\vec{\tau}_{n}\right)-\vec{B}\cdot\overleftrightarrow{g}\cdot\vec{s}_n
		\end{array}
		\right].
	\end{align}
	The isotropic AF Heisenberg interaction case corresponds to $\alpha=0$, and $\alpha=\pi/2$ indicates the $\Gamma_\perp$ only case. The direction of $\vec{B}$ is crucial for the energy spectrum, as we discuss in the following. We focus on the $\vec{B}\parallel\hat{x}$ case where analytical results can be derived. We also discuss the cases with $\vec{B}\perp\hat{x}$.

	\subsubsection{$\vec{B}\parallel\hat{x}$}

	For the magnetic field along the $x$ direction (the DM vector direction of the rung couplings), $\hat{H}_n$ [Eq.~(\ref{Eq:H_gen_rung})] can be diagonalized analytically. The energies are given by $E_1=-3J_0/4$, $E_2=J_0/4-g_xB_x$, $E_3=J_0/4$, and $E_2=J_0/4+g_xB_x$ as plotted in Fig.~\ref{Fig:General_rung_Bx}. The wavefunctions are given by
	\begin{subequations}\label{Eq:psi_generic_Bx}
		\begin{align}
			\psi_1=&\frac{1}{\sqrt{2}}\left[\cos\alpha(\ket{\uparrow_s\downarrow_{\tau}}-\ket{\downarrow_s\uparrow_{\tau}})-i\sin\alpha(\ket{\uparrow_s\uparrow_{\tau}}-\ket{\downarrow_s\downarrow_{\tau}})\right]\\
			\psi_2=&\frac{1}{2}\left(\ket{\uparrow_s\uparrow_{\tau}}+\ket{\uparrow_s\downarrow_{\tau}}+\ket{\downarrow_s\uparrow_{\tau}}+\ket{\downarrow_s\downarrow_{\tau}}\right)\\
			\psi_3=&\frac{1}{\sqrt{2}}\left[-i\sin\alpha(\ket{\uparrow_s\downarrow_{\tau}}-\ket{\downarrow_s\uparrow_{\tau}})+\cos\alpha(\ket{\uparrow_s\uparrow_{\tau}}-\ket{\downarrow_s\downarrow_{\tau}})\right]\\
			\psi_4=&\frac{1}{2}\left(\ket{\uparrow_s\uparrow_{\tau}}-\ket{\uparrow_s\downarrow_{\tau}}-\ket{\downarrow_s\uparrow_{\tau}}+\ket{\downarrow_s\downarrow_{\tau}}\right).
		\end{align}
	\end{subequations}
	Note that $\psi_1$ is chosen such that it is reduced to the well-known expression for the spin singlet state $\frac{1}{\sqrt{2}}(\ket{\uparrow_s\downarrow_{\tau}}-\ket{\downarrow_s\uparrow_{\tau}})$ at $\alpha=0$. For other values of $\alpha$, $\psi_1$ can be viewed as a generalized rung singlet state, as it does not respond to the $B_x$ magnetic field.
	
	The universal spectrum can be understood by the two-spin Hamiltonian in Eq.~(\ref{Eq:H_2spin_rotated}), in which the two-spin interactions are equivalent to an isotropic AF Heisenberg interaction. In our case, the local basis rotation does not modify the $B_x$ field. Thus, the energy spectrum of $\hat{H}_n$ under an $x$ directional magnetic field is universal and independent of $\alpha$. The wavefunctions still crucially depend on the value of $\alpha$, so does the effective spin model which we derive next.

	With the two low-energy states $\psi_1$ and $\psi_2$ given by Eq.~(\ref{Eq:psi_generic_Bx}) and the first-order perturbation given by Eq.~(\ref{Eq:Local_op}), the one-spin local operators in the reduced Hilbert space are given by
	\begin{subequations}
		\begin{align}
			s^x_n\rightarrow&\frac{1}{4}(I_n-\sigma_n^z),\\
			s^y_n\rightarrow&\frac{1}{2\sqrt{2}}\left(\sin\alpha\sigma_n^x+\cos\alpha\sigma_n^y\right),\\
			s^z_n\rightarrow&\frac{1}{2\sqrt{2}}\left(\cos\alpha\sigma_n^x-\sin\alpha\sigma_n^y\right),\\
			\tau^x_n\rightarrow&\frac{1}{4}(I_n-\sigma_n^z),\\
			\tau^y_n\rightarrow&\frac{1}{2\sqrt{2}}\left(\sin\alpha\sigma_n^x-\cos\alpha\sigma_n^y\right),\\
			\tau^z_n\rightarrow&\frac{1}{2\sqrt{2}}\left(-\cos\alpha\sigma_n^x-\sin\alpha\sigma_n^y\right).
		\end{align}
	\end{subequations}
		Next, we derive the contribution from $\hat{H}_{\text{leg}}$ [given by Eq.~(\ref{Eq:H_spin_ladder}a)] as follows:
		\begin{align}
			\nonumber&\hat{H}_{\text{leg}}\\
			\rightarrow\!
			&\sum_n\!\!\left\{\begin{array}{c}
				\frac{(J+J'-\Gamma-\Gamma')}{16}(I_n-\sigma_n^z)(I_{n+1}-\sigma_{n+1}^z)\\[1mm]
				+\frac{(J+J'-\Gamma-\Gamma')}{8}\left[\cos^2(\alpha)\sigma_n^x\sigma_{n+1}^x+\sin^2(\alpha)\sigma_n^y\sigma_{n+1}^y\right]\\[1mm]
				+\frac{(J+J'+\Gamma+\Gamma')}{8}\left[\sin^2(\alpha)\sigma_n^x\sigma_{n+1}^x+\cos^2(\alpha)\sigma_n^y\sigma_{n+1}^y\right]\\[1mm]
				+\frac{(\Gamma-\Gamma')}{4}\left[\cos(\alpha)\sin(\alpha)\left(\sigma_n^x\sigma_{n+1}^y+\sigma_n^y\sigma_{n+1}^x\right)\right]\\[1mm]
				\!\!+\frac{(D-D')}{8\sqrt{2}}\cos(\alpha)\!\left[\sigma_n^x(I_{n+1}-\sigma^z_{n+1})\!-\!(I_n-\sigma^z_n)\sigma_{n+1}^x\right]\!\!\\[1mm]
				\!\!-\frac{(D+D')}{8\sqrt{2}}\sin(\alpha)\!\left[\sigma_n^y(I_{n+1}-\sigma^z_{n+1})\!-\!(I_n-\sigma^z_n)\sigma_{n+1}^y\right]\!\!
			\end{array}
			\right\}.
		\end{align}

	Using the results above and ignoring the boundary terms, we can map the two-leg spin ladder into an effective spin-$1/2$ model (with $\mathcal{S}_n^{\mu}=\sigma_n^{\mu}/2$) given by
	\begin{align}\label{Eq:H_eff_general_Bx}
		\nonumber\hat{H}_{\text{eff}}=&\sum_n\left[\begin{array}{c}
			\tilde{J}_x\mathcal{S}_n^x\mathcal{S}_{n+1}^x+\tilde{J}_y\mathcal{S}_n^y\mathcal{S}_{n+1}^y+\tilde{J}_z\mathcal{S}_n^z\mathcal{S}_{n+1}^z\\[1mm]
			+\tilde{D}_x\left(\mathcal{S}_n^y\mathcal{S}_{n+1}^z-\mathcal{S}_n^z\mathcal{S}_{n+1}^y\right)\\[1mm]
			+\tilde{D}_y\left(\mathcal{S}_n^z\mathcal{S}_{n+1}^x-\mathcal{S}_n^x\mathcal{S}_{n+1}^z\right)\\[1mm]
			+\tilde\Omega_z\left(\mathcal{S}_n^x\mathcal{S}_{n+1}^y+\mathcal{S}_n^y\mathcal{S}_{n+1}^x\right)-\tilde{h}_z\mathcal{S}^z_n
		\end{array}
		\right]\\
		&+\frac{\tilde{J}_z}{2}\!\left(\mathcal{S}^z_{n_L}\!+\!\mathcal{S}^z_{n_R}\right)\!+\!\frac{\tilde{D}_y}{2}\!\left(\mathcal{S}^x_{n_L}\!-\!\mathcal{S}^x_{n_R}\right)\!-\!\frac{\tilde{D}_x}{2}\!\left(\mathcal{S}^y_{n_L}\!-\!\mathcal{S}^y_{n_R}\right),
	\end{align}
	where $n_R$ and $n_L$ are the right and left boundary sites,
	\begin{subequations}\label{Eq:couplings_Bx_gen}
		\begin{align}
			\tilde{J}_x=&(J+J')/2-(\Gamma+\Gamma')\cos(2\alpha)/2,\\
			\tilde{J}_y=&(J+J')/2+(\Gamma+\Gamma')\cos(2\alpha)/2,\\
			\tilde{J}_z=&(J+J'-\Gamma-\Gamma')/4,\\
			\tilde{D}_x=&(D+D')\sin(\alpha)/2\sqrt{2},\\
			\tilde{D}_y=&(D-D')\cos(\alpha)/2\sqrt{2},\\
			\tilde{\Omega}_x=&(\Gamma-\Gamma')\sin(2\alpha)/2,\\
			\tilde{h}_z=&J_\perp-g_xB_x+(J+J'-\Gamma-\Gamma')/4.
		\end{align}
	\end{subequations}
	The second line of Eq.~(\ref{Eq:H_eff_general_Bx}) describes the boundary terms.
	The effective spin-$1/2$ model is an XYZ spin chain with a DM interaction and external magnetic fields along both the $y$ and $z$ directions. In the limit $\alpha=0$, the result is reduced to Eq.~(\ref{Eq:H_eff_AF_Bx}). Notably, the symmetric anisotropic exchange $\tilde{\Omega}_x$ term, which is absent in the microscopic spin-spin interactions, arises as long as $\alpha\neq 0,\pi/2$ and $\Gamma\neq \Gamma'$. The DM interactions in the effective spin model cannot be completely turned off in generic situations that $0<\alpha<\pi/2$ and $D,D'\neq 0$.

	\subsubsection{$\vec{B}\perp\hat{x}$}

	\begin{figure}[t!]
		\includegraphics[width=0.45\textwidth]{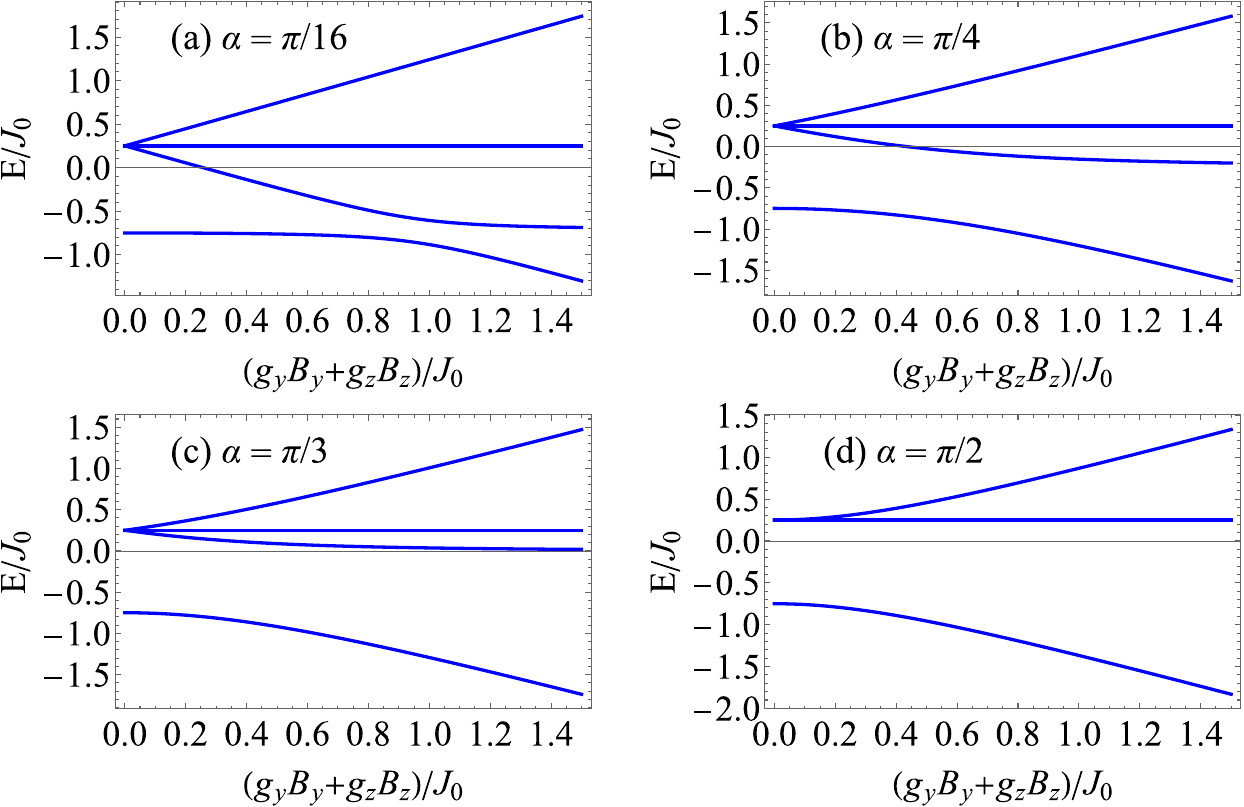}
		\caption{The energy spectrum of $\hat{H}_n$ [Eq.~(\ref{Eq:H_gen_rung})] under a magnetic field perpendicular to $\hat{x}$. (a) $\alpha=\pi/16$. (b) $\alpha=\pi/4$. (c) $\alpha=\pi/3$. (d) $\alpha=\pi/2$. Note that there are two degenerate energy levels that are independent of magnetic field in (d). The results show strong $\alpha$-dependence. The magnetic field induced level crossing is generically avoided for $\alpha\neq 0$.}
		\label{Fig:General_rung_By}
	\end{figure}
	
	For a magnetic field perpendicular to the $x$ direction, the energy spectrum depends crucially on $\alpha$ \cite{LiR2014} as shown in Fig.~\ref{Fig:General_rung_By}. Unlike the previous case with a magnetic field along the $x$ direction, we are unable to find analytical solutions even for the two-spin problem. Tuning slightly away from $\alpha=0$ [e.g., Fig.~\ref{Fig:General_rung_By}(a)], the energy spectrum is similar to Fig.~\ref{Fig:AF_rung} except that the level repulsion is present when the Zeeman energy is close to $J_0$. Such a level repulsion is due to the DM interaction, which generically takes place for $0<\alpha<\pi/2$. At $\alpha=\pi/2$, the DM interaction is absent, and $\hat{H}_{n}=J_0(s_n^x\tau_n^x-s_n^y\tau_n^y-s_n^z\tau_n^z)$. There are two degenerate excited states that are independent of the magnetic field strength as shown in Fig.~\ref{Fig:General_rung_By}(d). The gap between the ground state and the excited states increases as the magnetic field strength increases. 
	
	The strong $\alpha$ dependence of the spectrum puts constraints on the validity of the effective spin-$1/2$ model construction. For small $\alpha$'s, we can still derive effective spin-$1/2$ models as long as the two lower energy states are sufficiently separated from the higher two energy states. The results are qualitatively similar to the isotropic AF interaction limit with a magnetic field along $\hat{y}$ or $\hat{z}$ direction. Specifically, the DM interaction in the rung, the $D_\perp$ term, plays the role of a transverse field coupling in the effective model, which mixes the two states and tends to open up a gap, consistent with the level repulsion due to a finite $D_\perp$ ($\alpha\neq 0,\pi/2$) as shown in Fig.~\ref{Fig:General_rung_By}. When the value of $\alpha$ is sufficiently away from zero, the ground state is well separated from the excited energy levels. Thus, the low-energy phase of the spin ladder is mainly determined by the ground state of $\hat{H}_n$ [Eq.~(\ref{Eq:H_gen_rung})], and the perturbations from $\hat{H}_{\text{leg}}$ [given by Eq.~(\ref{Eq:H_spin_ladder}a)] only modify the results quantitatively.

	\subsection{Ground states of effective spin chains}
	
	Here, we discuss the ground states based on the strong rung coupling analysis. In the absence of a magnetic field, the ground state is a product state with each rung forming $\psi_1$. For a very large magnetic field, the ground state is another product state with each rung forming $\psi_2$, corresponding to a fully polarized state. These two cases correspond to a large $|\tilde{h}_z|$ in the effective spin chain, and the effective longitudinal magnetic field dictates the low-energy properties. For an intermediate magnetic field, the inter-rung correlation can build up, and one has to solve the effective spin$-1/2$ model. The general situations are very complicated, and numerical simulations (e.g., DMRG) are required to determine the precise phase diagram. In the following, we discuss a few situations focusing on the identical leg cases near the isotropic AF Heisenberg limit.
	
	First, we consider the cases with identical couplings of the two legs (i.e., $J=J'$, $D=D'$, and $\Gamma=\Gamma'$) in the isotropic AF rung coupling limit ($\alpha\rightarrow0$). For $\vec{B}\parallel\hat{x}$, the effective spin model [Eq.~(\ref{Eq:H_eff_AF_Bx})] is reduced to an XYZ chain (with $\tilde{J}_z<\tilde{J}_x<\tilde{J}_y$) under a $z$ directional magnetic field, and the ground states is characterized by a finite staggered magnetization along the $y$ direction for a small $|\tilde{h}_z|$ and a polarized state for a sufficiently large $|\tilde{h}_z|$. For $\vec{B}\parallel\hat{y}$, the effective model [Eq.~(\ref{Eq:H_eff_AF_By})] becomes an XXZ chain (with $\tilde{J}_z<\tilde{J}_{xy}$) with DM interactions and both $y$ and $z$ directional magnetic fields. This model has been studied in Refs.~\cite{GangadharaiahS2008,GarateI2010}, and states with finite staggered magnetization along the $x$ direction can be realized for a sufficiently large $\tilde{h}_y/\tilde{D}_z$. The detailed phase diagram for the XXZ chain with DM interaction and transverse magnetic field can be found in Ref.~\cite{GarateI2010}. For $\vec{B}\parallel\hat{x}$, the effective spin model [Eq.~(\ref{Eq:H_eff_AF_Bz})] is reduced to an XYZ model (with $\tilde{J}_z<\tilde{J}_x<\tilde{J}_y$) under both $y$ and $z$ directional magnetic fields. We expect fully polarized states along the $z$ ($y$) direction when the $\tilde{h}_z$ ($\tilde{h}_y$) term dominates. With small or negligible $\tilde{h}_y,\tilde{h}_z$, the ground state is described by a finite staggered magnetization along the $y$ direction.
	
	%As for the general rung coupling with $\vec{B}\parallel\hat{x}$, the effective spin model [Eq.~(\ref{Eq:H_eff_general_Bx})] is described by an XYZ chain $\tilde{J}_z<\tilde{J}_x<\tilde{J}_y$) with a DM interaction (DM vector along the $x$ direction) under a $z$ directional magnetic field. The resulting effective model can be reduced to the isotropic AF Heisenberg limit by setting $\alpha=0$. The new terms due to $\alpha\neq 0$ are the DM interaction $\tilde{D}_x$ and the symmetric anisotropic exchange ($\tilde{\Omega}_x$) interactions, which may play an important role for the phase diagram in the small $\tilde{h}_z$ regime. Numerical simulations are required to identify the complete phase diagram.

	Finally, we discuss the implications of ground states in the microscopic two-leg spin ladder model. The polarization along the $z$ ($-z$) direction corresponds to wavefunction $\psi_1$ ($\psi_2$) at each rung. The polarization perpendicular to the $z$ direction implies the linear combination of $\psi_1$ and $\psi_2$ at each rung. Owing to the inter-rung couplings, we also expect significant inter-rung correlation in the situations that are far away from polarized states in the effective spin model. For example, several cases may realize an XY phase with power-law correlation in the transverse spins. Such a phase can be described by a Luttinger liquid, and the ground state cannot be represented by a product state of individual rung wavefunctions.
	
	\subsection{Engineering DM interactions}\label{Sec:DM_engineering}

	The derived effective spin-$1/2$ models show concrete relations between the effective couplings and the microscopic interaction parameters. Particularly, the DM interaction, which naturally arises in quantum materials without inversion symmetry but is rarely explored in artificial quantum systems, can be engineered. We emphasize several important perspectives along this line.
	
	The DM interactions in the effective model generally manifest, but their parameter ranges are constrained by the microscopic relations discussed earlier in this section. In the spin qubit experiments, one can modify the electron tunnelings, both the spin-preserving and spin-flip processes, by the local electric gates and thereby tune the spin-spin interactions. We note that the DM vector cannot be tuned by gating, so $D,D',D_\perp>0$ in our consideration. Taking the isotropic AF case with $\vec{B}\parallel \hat{x}$ as an example, the effective DM interaction is given by $\tilde{D}_y=(D-D')/2\sqrt{2}$ [Eq.~(\ref{Eq:couplings_Bx_AF}d)], and $\tilde{D}_y$ can be tuned to positive, negative, or zero depending on the values of $D$ and $D'$. The same is true for the $\tilde{D}_y$ in the isotropic AF case with $\vec{B}\parallel \hat{z}$ [Eq.~(\ref{Eq:couplings_Bz_AF}d)]. However, this is not true for $\tilde{D}_z$ in Eq.~(\ref{Eq:couplings_By_AF}c) and $\tilde{D}_x$ in Eq.~(\ref{Eq:couplings_Bx_gen}d) because $\tilde{D}_z$ and $\tilde{D}_x$ depend on $D+D'$. The ability to turn off the DM interaction in the effective model is highly nontrivial, as the microscopic DM interactions are inevitable, i.e., $D,D'>0$ generically. Minimizing the DM interaction effects can allow for benchmark with the known Heisenberg interaction only results. 
	
	Among the quantum spin chain models with DM interactions, a particularly interesting case is the spin-$1/2$ chain with a staggered (i.e., alternating signs) DM interaction \cite{OshikawaM1997,AffleckI1999}. Under a uniform magnetic field, the staggered DM interaction effectively acts as a staggered field, resulting in a low-energy 1D sine-Gordon model description \cite{OshikawaM1997,AffleckI1999,OrignacE2007}. The staggered DM interaction can be engineered in the spin ladder model here by manipulating the microscopic interactions in the legs. Specifically, using $\tilde{D}_y=(D-D')/2\sqrt{2}$ in Eq.~(\ref{Eq:couplings_Bx_AF}d) (the isotropic AF case with $\vec{B}\parallel \hat{x}$), a staggered $\tilde{D}_y$ can be realized by requiring that $D>D'$ for the odd bonds and $D<D'$ for the even bonds. In this setup, a staggered transverse field term is also generated due to the local one-spin operator under projection, i.e., $s^x_n,\tau_n^x\rightarrow\left(I_n-\sigma_n^z\right)/4$. See Appendix~\ref{App:Staggered_DM} for a detailed discussion. Similarly, a staggered $\tilde{D}_y$ [Eq.~(\ref{Eq:couplings_Bz_AF}d)] can also be realized in the isotropic AF case with $\vec{B}\parallel \hat{z}$. Our results suggest that the Ge hole spin qubits can emulate models described by the 1D sine-Gordon theory, and the corresponding breather excitations of the sine-Gordon model, can be in principle investigated by the experiments in the spirit of the many-body electron spin resonance \cite{OshikawaM1999,OshikawaM2002}.

	The examples above show that the Ge hole quantum dot arrays are ideal for exploring quantum phases and dynamics of spin models with DM interactions, an unprecedented direction in synthetic quantum systems.

	\subsection{Boundary terms}
	
	Finally, we discuss the boundary terms in the effective spin-$1/2$ chain, i.e., the second lines of	Eqs.~(\ref{Eq:couplings_Bx_AF}), (\ref{Eq:couplings_By_AF}), (\ref{Eq:couplings_Bz_AF}), and (\ref{Eq:H_eff_general_Bx}). These boundary terms arise due to the missing spin-spin interactions along the legs (i.e., one bond on the boundary instead of two bonds in the bulk). Notably, these boundary terms manifest as local fields with the same longitudinal component ($z$) but opposite transverse components ($x$ and $y$). The longitudinal field modifies the precise transition point between $\psi_1$ and $\psi_2$ at the boundary rungs; the transverse field tends to enhance hybridization between $\psi_1$ and $\psi_2$. For small-size systems (e.g., the $2\times 4$ arrays in Refs.~\cite{ZhangX2025,FarinaPC2025a,BurkardG2023}), the existence of the boundary terms may play an important role in the phases and the corresponding characterization.

	\section{Weak rung coupling limit}\label{Sec:Weak_rung}

	In the weak rung coupling limit, we analyze the ground state of each individual leg (chain) and then construct the phase diagram for the two-leg spin ladder, treating rung couplings as perturbations. For simplicity, we study the identical legs case. In the absence of the rung couplings, each leg can be mapped into an isotropic AF Heisenberg chain after local basis rotations as discussed in Sec.~\ref{Sec:Model}. Thus, the two-leg spin ladder problem is equivalent to coupled AF Heisenberg chains, which can be studied using field theoretical methods. Our main goal is to provide phase diagrams in several analytically tractable limits.
	
	In the remainder of this section, we first discuss the basis rotations that are convenient for the bosonization analysis. Then, we employ Abelian bosonization \cite{GogolinAO2004,GiamarchiT2003} for the problem with a longitudinal magnetic field  ($\vec{B}\parallel\hat{y}$), uncovering the effect of the DM interactions and constructing low-energy phase diagrams. We also discuss the phase diagram for the transverse field case ($\vec{B}\perp\hat{y}$).\\

	\subsection{Basis rotation}
	
	In the weak rung coupling limit, we focus on $\hat{H}_{\text{leg}}$ [given by Eq.~(\ref{Eq:H_spin_ladder}a)] and analyze the properties exactly. As we discussed in Sec.~\ref{Sec:Model}, the DM interactions in each leg can be gauged away by a local basis rotation, encoded by a site-dependent rotation operator $\hat{U}$ such that
	\begin{subequations}
		\begin{align}
			\hat{U}s^x_n\hat{U}^{-1}=&s^x_n\cos(n\eta_s)-s^z_n\sin(n\eta_s),\\
			\hat{U}s^y_n\hat{U}^{-1}=&s^y_n,\\
			\hat{U}s^z_n\hat{U}^{-1}=&s^z_n\cos(n\eta_s)+s^x_n\sin(n\eta_s),\\
			\hat{U}\tau^x_n\hat{U}^{-1}=&\tau^x_n\cos(n\eta_\tau)-\tau^z_n\sin(n\eta_\tau),\\
			\hat{U}\tau^y_n\hat{U}^{-1}=&\tau^y_n,\\
			\hat{U}\tau^z_n\hat{U}^{-1}=&\tau^z_n\cos(n\eta_\tau)+\tau^x_n\sin(n\eta_\tau),
		\end{align}
	\end{subequations}
	where $\eta_s=\tan^{-1}[D/(J-\Gamma)]$ and $\eta_\tau=\tan^{-1}[D'/(J'-\Gamma')]$. After the local basis rotation, the spin-ladder Hamiltonian becomes $\hat{H}'_{\text{2-leg}}=\hat{U}\hat{H}_{\text{2-leg}}\hat{U}^{-1}=\hat{H}_{\text{leg}}'+\hat{H}_{\text{rung}}'+\hat{H}_{B}'$, where 
	\begin{widetext}
		\begin{subequations}\label{Eq:H_spin_ladder_rotd}
			\begin{align}
			\hat{H}_{\text{leg}}'=&\sum_n\left[\begin{array}{c}
					(J+\Gamma)s^y_{n}s^y_{n+1}+\sqrt{(J-\Gamma)^2+D^2}\left(s^x_{n}s^x_{n+1}+s^z_{n}s^z_{n+1}\right)\\[1mm]
					+(J'+\Gamma')\tau^y_{n}\tau^y_{n+1}+\sqrt{(J'-\Gamma')^2+D'^2}\left(\tau^x_{n}\tau^x_{n+1}+\tau^z_{n}\tau^z_{n+1}\right)
				\end{array}\right],\\
				\hat{H}_{\text{rung}}'=&\sum_n\left\{\begin{array}{c}
					(J_{\perp}-\Gamma_\perp)\left[s_n^y\tau_n^y+\left(s_n^x\tau_n^x+s_n^z\tau_n^z\right)\cos\left(n(\eta_s-\eta_\tau)\right)+\left(s_n^x\tau_n^z-s_n^z\tau_n^x\right)\sin\left(n(\eta_s-\eta_\tau)\right)\right]\\[1mm]
					+2\Gamma_\perp\left[\begin{array}{c}
						s_n^x\tau_n^x\cos\left(n\eta_s\right)\cos\left(n\eta_\tau\right)+s_n^z\tau_n^z\sin\left(n\eta_s\right)\sin\left(n\eta_\tau\right)\\[1mm]
						-s_n^x\tau_n^z\cos\left(n\eta_s\right)\sin\left(n\eta_\tau\right)-s_n^z\tau_n^x\sin\left(n\eta_s\right)\cos\left(n\eta_\tau\right)
					\end{array}\right]\\[3mm]
					+D_{\perp}\left[s^y_n\tau^z_n\cos(n\eta_\tau)+s^y_n\tau^x_n\sin(n\eta_\tau)-s^z_n\tau_n^y\cos(n\eta_s)-s^x_n\tau_n^y\sin(n\eta_s)\right]
				\end{array}
				\right\},\\[2mm]
				\hat{H}_{B}'=&-\sum_n\left\{\begin{array}{c}
					g_xB_x\left[s^x_n\cos(n\eta_s)-s^z_n\sin(n\eta_s)+\tau^x_n\cos(n\eta_\tau)-\tau^z_n\sin(n\eta_\tau)\right]
					+g_yB_y\left[s^y_n+\tau^y_n\right]\\[1mm]
					+g_zB_z\left[s^z_n\cos(n\eta_s)+s^x_n\sin(n\eta_s)+\tau^z_n\cos(n\eta_\tau)+\tau^x_n\sin(n\eta_\tau)\right]
				\end{array}
				\right\}.
			\end{align}
		\end{subequations}
	\end{widetext}
	The local basis rotation results in a simplified $\hat{H}_{\text{leg}}'$, providing a good starting point for weak rung coupling analysis. The price we pay here is that $\hat{H}_{\text{rung}}'$ and part of $\hat{H}_{B}'$ (i.e., $B_x$ and $B_z$ terms) become quite complicated. 
	
	A particularly interesting limit is the identical legs limit, i.e., $J=J'$, $D=D'$, and $\Gamma=\Gamma'$. In this section, we focus mainly on such a special limit as the couplings in $\hat{H}_{\text{rung}}'$ are greatly simplified. For computational convenience, we further perform a global change of basis, $s_n^y\rightarrow s_n^z$, $s_n^z\rightarrow -s_n^y$, $\tau_n^y\rightarrow \tau_n^z$, and $\tau_n^z\rightarrow -\tau_n^y$, so that the longitudinal direction of $\hat{H}_{\text{leg}}'$ coincides with $\hat{z}$ in the rotated basis. After the simplification and the basis changes, $\hat{H}'_{\text{2-leg}}\rightarrow \hat{H}_0+\hat{H}_{\perp,1}+\hat{H}_{\perp,2}+\hat{H}_{B_{xz}}$, where
	\begin{widetext}
		\begin{subequations}\label{Eq:H_0_H_perp_H_Bxz}
			\begin{align}
				\hat{H}_0=&\sum_n\left[\begin{array}{c}
					\frac{1}{2}\sqrt{(J-\Gamma)^2+D^2}\left(s^+_{n}s^-_{n+1}+s^-_{n}s^+_{n+1}\right)+(J+\Gamma)s^z_{n}s^z_{n+1}-g_yB_ys_n^z\\[1mm]
					+\frac{1}{2}\sqrt{(J-\Gamma)^2+D^2}\left(\tau^+_{n}\tau^-_{n+1}+\tau^-_{n}\tau^+_{n+1}\right)+(J+\Gamma)\tau^z_{n}\tau^z_{n+1}-g_yB_y\tau_n^z
				\end{array}\right],\\
				\hat{H}_{\perp,1}
				=&\sum_n\left[
					(J_{\perp}-\Gamma_\perp)s_n^z\tau_n^z
					+\frac{J_\perp}{2}\left(s_n^+\tau_n^-+s_n^-\tau_n^+\right)
					+\frac{\Gamma_\perp}{2}\left(s_n^+\tau_n^+e^{-i2n\eta}+s_n^-\tau_n^-e^{i2n\eta}\right)
				\right]\\
				\hat{H}_{\perp,2}
				=&\frac{D_{\perp}}{2i}\sum_n
					\left[s^z_n\left(-\tau_n^+e^{-in\eta}+\tau_n^-e^{in\eta}\right)-\left(-s_n^+e^{-in\eta}+s_n^-e^{in\eta}\right)\tau^z_n\right]\\
				\hat{H}_{B_{xz}}
				=&-\sum_n\left\{\begin{array}{c}
					\frac{g_xB_x}{2}\left[s^+_ne^{-in\eta}+s^-_ne^{in\eta}+\tau^+_ne^{-in\eta}+\tau^-_ne^{in\eta}\right]\\[1mm]
					+\frac{g_zB_z}{2i}\left[-s^+_ne^{-in\eta}+s^-_ne^{in\eta}-\tau^+_ne^{-in\eta}+\tau^-_ne^{in\eta}\right]
				\end{array}
				\right\}.
			\end{align}
		\end{subequations}
	\end{widetext}
	In the above expressions, $s_n^{\pm}\equiv s_n^x\pm is_n^y$, $\tau_n^{\pm}\equiv \tau_n^x\pm i\tau_n^y$, and $\eta=\tan^{-1}[D/(J-\Gamma)]$. $\hat{H}_0$ describes two identical XXZ spin-$1/2$ chains under the longitudinal magnetic field, which can be mapped to Luttinger liquids \cite{GogolinAO2004,GiamarchiT2003}; $\hat{H}_{\perp,1}$ and $\hat{H}_{\perp,2}$ are the rung couplings (i.e., the inter-chain couplings); $\hat{H}_{B_{xz}}$ describes the magnetic field coupling in the physical $xz$ plane. The phase factor $n\eta$ appears whenever the interaction breaks the $U(1)$ longitudinal spin conservation. Note that we have performed several basis rotations for the spins while keeping the coupling constants $g_xB_x$, $g_yB_y$, and $g_zB_z$ unchanged, reflecting with the physical basis. 
	
	In the absence of $D_\perp$ and $\Gamma_\perp$, the rung coupling is reduced to the isotropic AF Heisenberg interaction, and the results are the same as the known AF Heisenberg spin ladder models \cite{GogolinAO2004,GiamarchiT2003}. Our goals are to uncover the unprecedented phases driven by $D_\perp$, $\Gamma_\perp$, and $\eta$.

	\subsection{Bosonization}

	A powerful way to analyze the low-temperature phase diagrams of 1D systems is Abelian bosonization \cite{GiamarchiT2003,GogolinAO2004}. First, we perform Jordan-Wigner transformation and then take the continuum limit \cite{GiamarchiT2003}. Concomitantly, we define the spin operators in the continuum, $s^{\pm}(x)|_{x=na}=s^{\pm}_n/\sqrt{a}$, $s^{z}(x)|_{x=na}=s^{z}_n/a$, $\tau^{\pm}(x)|_{x=na}=\tau^{\pm}_n/\sqrt{a}$, and $\tau^{z}(x)|_{x=na}=\tau^{z}_n/a$, where $a$ is the lattice constant along the leg direction. 
	Then, we employ the standard Abelian bosonization for spin-$1/2$ operators \cite{OrignacE1997,OrignacE1998,GiamarchiT2003}. The spin operators are bosonized to
	\begin{subequations}\label{Eq:Bosonization_s_tau}
		\begin{align}
			s^z(x)\rightarrow&\frac{1}{\pi}\partial_x\theta_1(x)+\frac{1}{\pi a}(-1)^{x/a}\cos\left[2\theta_1(x)\right],\\
			s^{\pm}(x)\rightarrow&\frac{1}{\sqrt{2\pi a}}e^{\pm i\phi_1(x)}\!\!\left\{(-1)^{x/a}+\cos\left[2\theta_1(x)\right]\right\},\\
			\tau^z(x)\rightarrow&\frac{1}{\pi}\partial_x\theta_2(x)+\frac{1}{\pi a}(-1)^{x/a}\cos\left[2\theta_2(x)\right],\\
			\tau^{\pm}(x)\rightarrow&\frac{1}{\sqrt{2\pi a}}e^{\pm i\phi_2(x)}\!\!\left\{(-1)^{x/a}+\cos\left[2\theta_2(x)\right]\right\},
		\end{align}
	\end{subequations}
	where $\theta_1$ and $\theta_2$ are the bosonic fields corresponding to the longitudinal spin density, $\phi_1$ and $\phi_2$ are the bosonic fields corresponding to the phases, and $k_F=\pi/(2a)$ is used in the above expressions, i.e., assuming zero magnetization $\langle s_n^z\rangle=\langle \tau_n^z\rangle=0$. We have used the lattice constant $a$ for the ultraviolet length scale. Here, the bosonic fields satisfy $[\phi_n(x),\theta_{n'}(y)]=-i\delta_{n,n'}\pi \sgn(y-x)$ with $\sgn(z)$ being the sign function. The Klein factors are ignored as they are not essential to this work.
	
	\begin{widetext}
		Taking the continuum limit and using Eq.~(\ref{Eq:Bosonization_s_tau}), the Hamiltonians in Eq.~(\ref{Eq:H_0_H_perp_H_Bxz}) are expressed by
		\begin{subequations}\label{Eq:H_bosonized_12_original}
			\begin{align}
				\hat{H}_{0}\rightarrow&\hat{H}_{0,b}=\sum_{j=1,2}\int dx
				\left\{\frac{v}{2\pi}\left[K\left(\partial_x\phi_j\right)^2+\frac{1}{K}\left(\partial_x\theta_j\right)^2\right]-\frac{(J+\Gamma)}{2\pi^2a}\cos\left(4\theta_j\right)-g_yB_y\frac{\partial_x\theta_j}{\pi}\right\},\\
				\hat{H}_{\perp,1}\rightarrow&\hat{H}_{\perp,1,b}=\int dx\left\{\begin{array}{c}
					(J_\perp-\Gamma_{\perp})\left[\frac{a}{\pi^2}\partial_x\theta_1\partial_x\theta_2+\frac{1}{2\pi^2 a}\cos\left(2\theta_1+2\theta_2\right)+\frac{1}{2\pi^2 a}\cos\left(2\theta_1-2\theta_2\right)\right]\\[2mm]
					+\frac{J_\perp}{2\pi a}\cos\left(\phi_1-\phi_2\right)
					+\frac{\Gamma_\perp}{2\pi a}\cos\left(\phi_1+\phi_2-2\eta x/a\right)
					\end{array}
				\right\},\\
				\hat{H}_{\perp,2}\rightarrow&\hat{H}_{\perp,2,b}=\int dx\left\{\begin{array}{c}
					\frac{D_\perp}{\sqrt{2\pi}}\left[\left(\partial_x\theta_2\right)\sin\left(\phi_1-\eta x/a\right)\cos\left(2\theta_1\right)-\left(\partial_x\theta_1\right)\sin\left(\phi_2-\eta x/a\right)\cos\left(2\theta_2\right)
					\right]\\[2mm]
					+\frac{D_\perp}{\pi\sqrt{2\pi} a}\left[\sin\left(\phi_1-\eta x/a\right)\cos\left(2\theta_2\right)-\sin\left(\phi_2-\eta x/a\right)\cos\left(2\theta_1\right)
					\right]
				\end{array}
				\right\},\\
				\hat{H}_{B_{xz}}\rightarrow&\hat{H}_{B_{xz},b}=\int dx \left\{\begin{array}{c}
					-\frac{g_xB_x}{\sqrt{2\pi }a}\left[\cos\left(\phi_1-\eta x/a\right)\cos\left(2\theta_1\right)+\cos\left(\phi_2-\eta x/a\right)\cos\left(2\theta_2\right)\right]\\[2mm]
					-\frac{g_zB_z}{\sqrt{2\pi }a}\left[\sin\left(\phi_1-\eta x/a\right)\cos\left(2\theta_1\right)+\sin\left(\phi_2-\eta x/a\right)\cos\left(2\theta_2\right)\right]
				\end{array}
				\right\},
			\end{align}
		\end{subequations}
	\end{widetext}
		where $v$ is the velocity, and $K$ is the Luttinger parameter within each chain. Several bosonized expressions for the rung interactions are summarized in Appendix~\ref{App:Bosonized_interaction}. We have ignored the highly oscillatory $(-1)^{x/a}$ terms and several subleading terms in $\hat{H}_{\perp,b}$, and the results in this work do not depend on these omitted terms. The precise values of $v$ and $K$ can be extracted via the Bethe ansatz \cite{GiamarchiT2003,GiamarchiT1999}. In our case, the microscopic spin-spin interaction in the leg is close to the isotropic AF Heisenberg interaction as discussed in Sec.~\ref{Sec:Model}. Therefore, we conclude that $K=1/2$, and the low-energy phase of each leg is described by a gapless phase with the marginal interaction $\cos(4\theta_j)$  \cite{GiamarchiT2003}.

		Finally, we discuss $\hat{H}_{\perp,2,b}$ and $\hat{H}_{B_{xz},b}$, which break the longitudinal spin $U(1)$ symmetry and possess finite conformal spins. In the weak rung coupling limit, the dominant contribution comes from the second order of the transverse field (e.g., $B_x^2$, $B_z^2$), in which a new perturbation term $\cos(2\phi_j-2\eta x/a)$ is generated under RG \cite{GangadharaiahS2008,GarateI2010,NersesyanAA1993a,YakovenkoVM1992,GogolinAO2004}. The other second-order contributions (e.g., $D_\perp^2$, $D_\perp B_x$, and $D_\perp B_z$) can renormalize the existing terms in $\hat{H}_{0,b}$ and $\hat{H}_{\perp,1,b}$ and generate terms similar to $\cos(2\phi_j-2\eta x/a)$. As we will show later, the instabilities driven by the $U(1)$ preserving rung couplings are at the first order of couplings, so that we can ignore the $D_\perp^2$ contributions for the generic situations. The transverse field is also crucial for the low-energy phase diagram. One can construct the phase diagram based on the known results of the single chain model in Refs.~\cite{GangadharaiahS2008,GarateI2010}, as we will discuss at the end of this section. 
		
		\subsection{Symmetric and antisymmetric variables}
				
		Before proceeding further, we introduce a change of variables that is convenient for constructing the low-energy phase diagram. The Hamiltonian $\hat{H}_{0,b}+\hat{H}_{\perp,1,b}$ can be expressed in terms of the chain-symmetric and chain-antisymmetric variables, which are defined as:
		\begin{subequations}
			\begin{align}
				\Theta_{\pm}=\frac{1}{\sqrt{2}}\left(\theta_1\pm \theta_2\right),\,\,\,
				\Phi_{\pm}=\frac{1}{\sqrt{2}}\left(\phi_1\pm\phi_2\right).
			\end{align}
		\end{subequations}
		The Hamiltonian $\hat{H}=\hat{H}_{0,b}+\hat{H}_{\perp,1,b}$ can be reorganized into $\hat{H}_{+}+\hat{H}_{-}+\hat{H}_{\text{mixing}}$, where
		\begin{widetext}
			\begin{subequations}\label{Eq:H_+-}
				\begin{align}
					\hat{H}_+=&\int dx
					\left\{\frac{v_+}{2\pi}\left[K_+\left(\partial_x\Phi_+\right)^2+\frac{1}{K_+}\left(\partial_x\Theta_+\right)^2\right]+\frac{U_+}{2\pi a}\cos\left(2\sqrt{2}\Theta_+\right)+\frac{V_+}{2\pi a}\cos\left(\sqrt{2}\Phi_+-2\eta x/a\right)-\frac{g_yB_y\sqrt{2}}{\pi}\partial_x\Theta_+\right\},\\
					\hat{H}_-=&\int dx
					\left\{\frac{v_-}{2\pi}\left[K_-\left(\partial_x\Phi_-\right)^2+\frac{1}{K_-}\left(\partial_x\Theta_-\right)^2\right]+\frac{U_-}{2\pi a}\cos\left(2\sqrt{2}\Theta_-\right)+\frac{V_-}{2\pi a}\cos\left(\sqrt{2}\Phi_-\right)\right\},\\
					\hat{H}_{\text{mixing}}=&-\frac{U}{\pi a}\int dx\cos\left(2\sqrt{2}\Theta_+\right)\cos\left(2\sqrt{2}\Theta_-\right).
				\end{align}
			\end{subequations}
		\end{widetext}
		In the above expressions, $v_{\pm}$ and $K_{\pm}$ are the velocity and Luttinger parameter, which are given by
		\begin{align}
			v_\pm=v\sqrt{1\pm\frac{K(J_\perp-\Gamma_\perp)a}{\pi v}},\,\, K_\pm=\frac{K}{\sqrt{1\pm\frac{K(J_\perp-\Gamma_\perp)a}{\pi v}}}.
		\end{align}
		The hierarchy of $K_+$ and $K_-$ depends on $J_\perp-\Gamma_\perp$. For $J_\perp>\Gamma_\perp$ ($J_\perp<\Gamma_\perp$), one can show that $K_+<1/2<K_-$ ($K_-<1/2<K_+$). In the weak rung coupling limit, we expect that $K_+$ and $K_-$ are very close to the single-chain Luttinger parameter $K=1/2$. The coupling constants for the cosine terms in Eq.~(\ref{Eq:H_+-}) are given by $U_+=U_-=(J_\perp-\Gamma_\perp)/\pi$, $V_+=\Gamma_+$, $V_-=J_\perp$, and $U=(J+\Gamma)/\pi$. For $B_y=0$ and $\eta=0$, the leading RG flows for the interaction terms (based on the scaling dimensions) are given by
		\begin{subequations}
			\begin{align}
				\frac{dU_+}{dl}=&\left(2-2K_+\right)U_+,\,\,\,
				\frac{dV_+}{dl}=\left(2-\frac{1}{2K_+}\right)V_+,\\
				\frac{dU_-}{dl}=&\left(2-2K_-\right)U_-,\,\,\,
				\frac{dV_-}{dl}=\left(2-\frac{1}{2K_-}\right)U_-,\\
				\frac{dU}{dl}=&\left(2-2K_+-2K_-\right)U.
			\end{align}
		\end{subequations}
		The $U_\pm$ and $V_{\pm}$ terms are relevant at $K_\pm=1/2$. We can safely ignore the subleading $U$ term, which is less relevant. Next, we discuss the effects of finite $B_y$ and $\eta$, which are crucial for the low-energy phase diagram. We note that $\hat{H}_{\perp,2,b}$ [Eq.~(\ref{Eq:H_bosonized_12_original})] can contribute to the above RG flows at the order of $D_\perp^2$, which are negligible in the weak rung coupling limit.
		
		The $B_y$ term in $\hat{H}_+$ [Eq.~(\ref{Eq:H_+-}a)] can be eliminated by redefinition of $\Theta$. Particularly, we define $\Theta_+'= \Theta_+-\sqrt{2}\zeta x$, where $\zeta=K_+g_yB_y/(\pi v_+)$. With the $\Theta_+'$ variable, the $B_y$ term is eliminated, and the $U_+$ term becomes
		\begin{align}
			\frac{U_+}{2\pi a}\cos\left(2\sqrt{2}\Theta_+\right)=\frac{U_+}{2\pi a}\cos\left(2\sqrt{2}\Theta_+'+4\zeta x\right).
		\end{align}
		The situation here is equivalent to the Pokrovsky-Talapov commensurate-incommensurate problem \cite{PokrovskyVL1979}. A sufficiently large $\zeta$ (related to the applied longitudinal field $B_y$) drives the system into an incommensurate phase, making $U_+$ irrelevant. Similarly, there is a commensurate-incommensurate transition in the $V_+$ term tuned by $\eta/a$, which reflects the DM interaction along the chain. The phases here are determined by the interplay of $U_\pm$, $V_\pm$, $\zeta$, and $\eta$.

		\subsection{Order parameters}\label{SubSec:Order_parameter}
		
		To understand the ground state properties, we introduce the order parameters in this subsection. For the spin ladder problem, it is natural to define chain-symmetric and chain-antisymmetric spin order parameters. In addition, we separate the uniform and the staggered parts of spin configurations. Taking $s^z$ as an example [Eq.~(\ref{Eq:Bosonization_s_tau})], the uniform part is $\partial_x\theta/\pi$, and the staggered part is $\cos(2\theta_1)/(\pi a)$. One can use the factor $(-1)^{x/a}$ to filter out the staggered part. The pertinent order parameters are defined as follows:
		\begin{widetext}
			\begin{subequations}\label{Eq:Order_para}
				\begin{align}
					 m^Y_{\pm}(x)=&\frac{1}{\sqrt{2}}\left[s^z(x)\pm\tau^z(x)\right]_{\text{uniform}}\rightarrow \frac{1}{\pi}\partial_x\Theta_{\pm},\\
					 m^X_\pm(x)=&\frac{1}{\sqrt{2}}\left[s^x(x)\pm\tau^x(x)\right]_{\text{uniform}}\rightarrow\frac{1}{2\sqrt{\pi a}}\left[\cos(\phi_1)\cos(2\theta_1)\pm\cos(\phi_2)\cos(2\theta_2)\right],\\
					m^{Z}_{\pm}(x)=&\frac{1}{\sqrt{2}}\left[s^y(x)\pm\tau^y(x)\right]_{\text{uniform}}\rightarrow\frac{1}{2\sqrt{\pi a}}\left[\sin(\phi_1)\cos(2\theta_1)\pm\sin(\phi_2)\cos(2\theta_2)\right],\\
					n^{Y}_{+}(x)=&\frac{1}{\sqrt{2}}\left[s^z(x)+\tau^z(x)\right]_{\text{staggered}}\rightarrow \frac{\sqrt{2}}{\pi a}\cos(\sqrt{2}\Theta_+)\cos(\sqrt{2}\Theta_-),\\
					n^{Y}_{-}(x)=&\frac{1}{\sqrt{2}}\left[s^z(x)-\tau^z(x)\right]_{\text{staggered}}\rightarrow-\frac{\sqrt{2}}{\pi a}\sin(\sqrt{2}\Theta_+)\sin(\sqrt{2}\Theta_-),\\
					n^{X}_{+}(x)=&\frac{1}{\sqrt{2}}\left[s^x(x)+\tau^x(x)\right]_{\text{staggered}}\rightarrow\frac{1}{\sqrt{\pi a}}\cos(\Phi_+/\sqrt{2})\cos(\Phi_-/\sqrt{2}),\\
					n^{X}_{-}(x)=&\frac{1}{\sqrt{2}}\left[s^x(x)-\tau^x(x)\right]_{\text{staggered}}\rightarrow-\frac{1}{\sqrt{\pi a}}\sin(\Phi_+/\sqrt{2})\sin(\Phi_-/\sqrt{2}),\\
					n^{Z}_{+}(x)=&\frac{1}{\sqrt{2}}\left[s^y(x)+\tau^y(x)\right]_{\text{staggered}}\rightarrow\frac{1}{\sqrt{\pi a}}\sin(\Phi_+/\sqrt{2})\cos(\Phi_-/\sqrt{2}),\\
					n^{Z}_{-}(x)=&\frac{1}{\sqrt{2}}\left[s^y(x)-\tau^y(x)\right]_{\text{staggered}}\rightarrow\frac{1}{\sqrt{\pi a}}\cos(\Phi_+/\sqrt{2})\sin(\Phi_-/\sqrt{2}),
				\end{align}
			\end{subequations}
		\end{widetext}
		 where $m_\pm^{X,Y,Z}$ denotes for the uniform magnetization density, and $n_\pm^{X,Y,Z}$ denotes the staggered magnetization density. We note that the capitalized superscripts $X$, $Y$, and $Z$ correspond to the physical coordinates (i.e., coordinates defined in Fig.~\ref{Fig:SL}), not the rotated spin components (lower-case superscripts in $s$ and $\tau$) that are convenient for utilizing bosonization. Most of the order parameters have simplified expressions in terms of $\Theta_\pm$ and $\Phi_\pm$, except for $m^X_\pm$ and $m^Z_\pm$.

		 For gapped states, the order parameters can have nonzero expectation values, which describe the underlying ground-state spin textures. For gapless states, several order parameters show power-law correlation, and the dominant order is determined by the slowest varying order parameter \cite{GiamarchiT2003}, which corresponds to the most divergent susceptibility of the associated order. The dominant power-law correlation is expected to be stabilized in a finite system.

		\subsection{Phase diagram with a longitudinal magnetic field}

		\begin{figure}[t!]
			\includegraphics[width=0.5\textwidth]{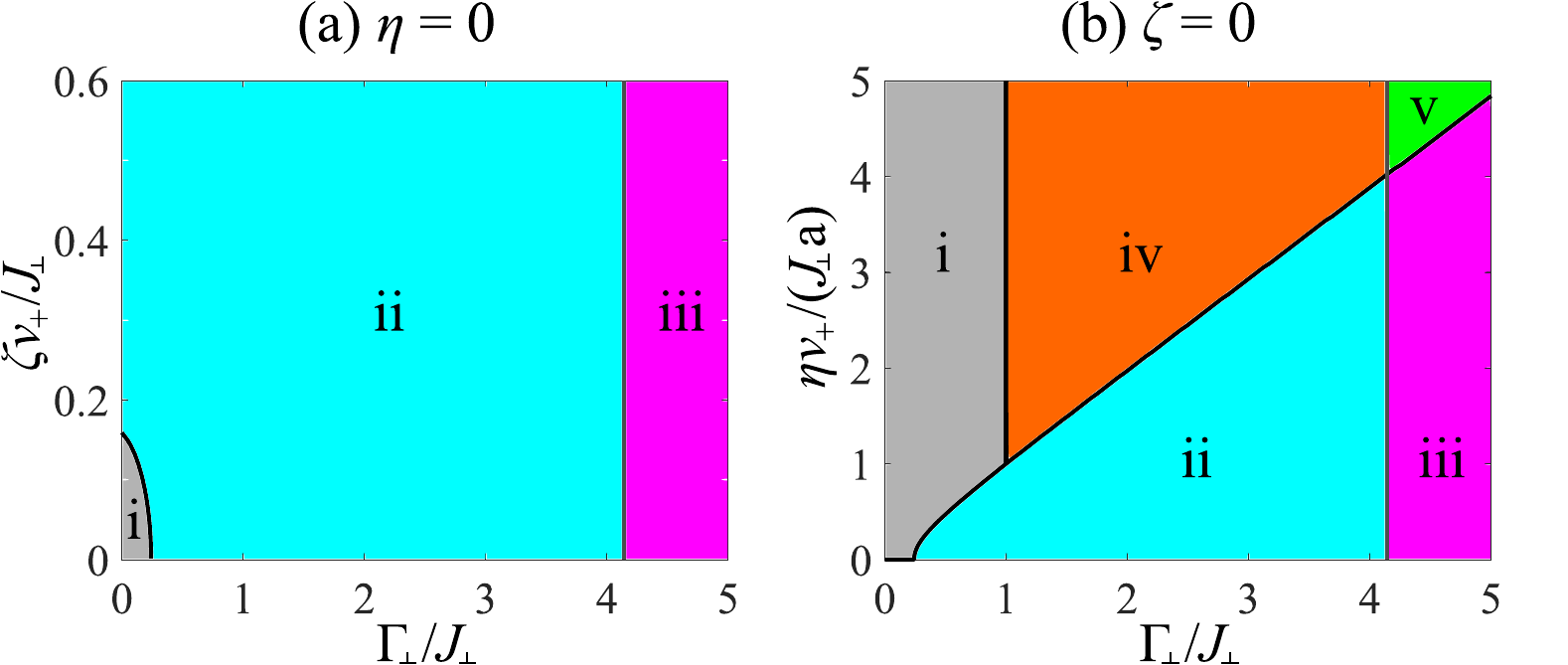}
			\caption{Spin ladder phase diagrams with $\eta=0$ or $\zeta=0$. (a) $\eta=0$, corresponding to zero DM interaction in the legs. Phase i is driven into phase ii for a sufficiently large $\zeta$. (b) $\zeta=0$, corresponding to zero magnetic field. Phase ii transits into phase i or iv for a sufficiently large $\eta$; phase iii is replaced by phase v in the large $\eta$ limit. The phases are summarized in Table~\ref{tab:phases}. See main text for detailed discussions.}
			\label{Fig:Ladder_PD_zero}
		\end{figure}

		\begin{figure}[t]
			\includegraphics[width=0.275\textwidth]{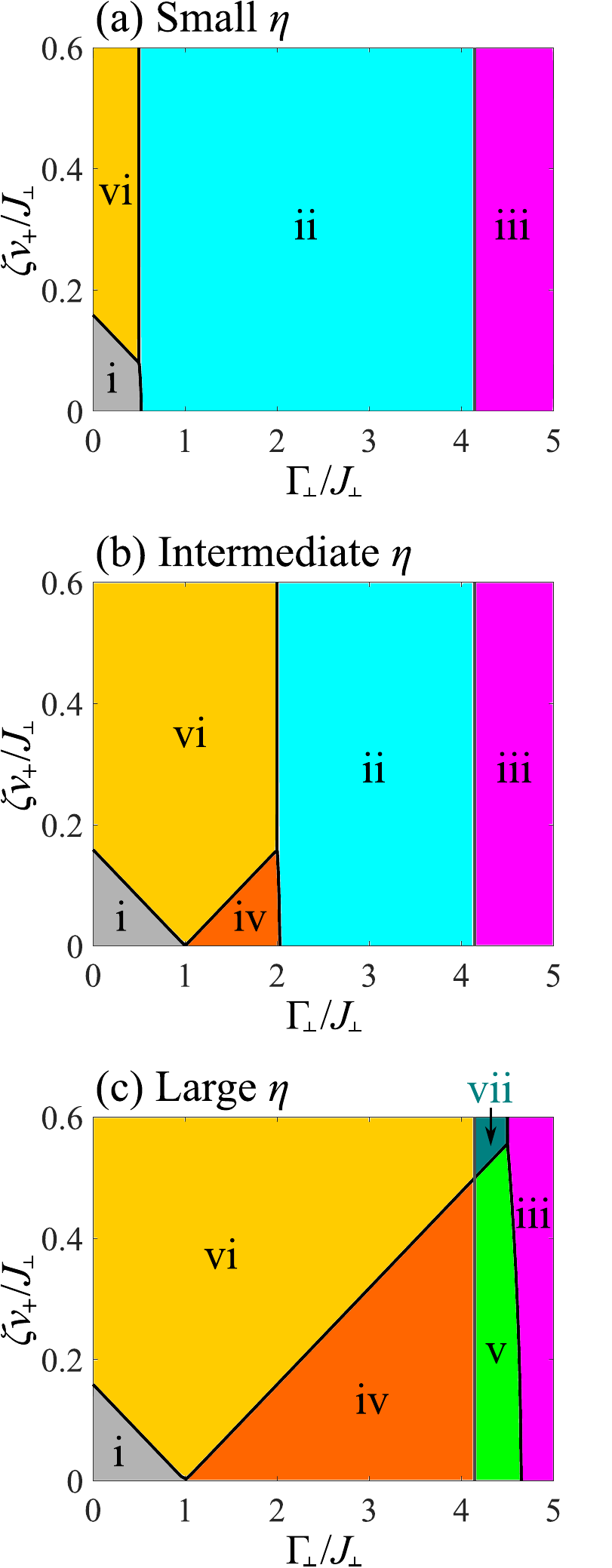}
			\caption{Spin ladder phase diagrams as functions of $\Gamma_\perp/J_\perp$ and $\zeta$ (associated with the longitudinal magnetic field) for different values of $\eta$ (rotated angle due to DM interactions in the legs). The phase boundaries are based on Luther-Emery theory discussed in Appendix~\ref{App:LE}. We use $\eta v_+/(J_\perp a)=$0.5, 2, and 4.5 for (a) Small $\eta$, (b) Intermediate $\eta$, and (c) Large $\eta$ cases, respectively. All the phases are summarized in Table~\ref{tab:phases}. See main text for detailed discussions.
			}
			\label{Fig:Ladder_PD_By}
		\end{figure}

		\begin{table*}[t]
			\centering
			\begin{tabular}{c||c|c|c|c|c}
				\toprule
				Phase & Constraints in $\Theta_+,\Phi_+$ & Constraints in $\Theta_-,\Phi_-$ & Leading Order &Subleading Order & Excitation \\
				\hline
				i & $\langle\cos(2\sqrt{2}\Theta_+)\rangle=-1$ & $\langle\cos(\sqrt{2}\Phi_-)\rangle=-1$ &  None & None & Gapped \\
				\hline
				ii & $\langle\cos(\sqrt{2}\Phi_+)\rangle=-1$ & $\langle\cos(\sqrt{2}\Phi_-)\rangle=-1$ & $\langle n_-^X\rangle$ & None & Gapped\\
				\hline
				iii & $\langle\cos(\sqrt{2}\Phi_+)\rangle=-1$ & $\langle\cos(2\sqrt{2}\Theta_-)\rangle=1$ & None & None & Gapped \\
				\hline
				iv & $\langle\cos(2\sqrt{2}\Theta_+)\rangle=1$ & $\langle\cos(\sqrt{2}\Phi_-)\rangle=-1$ & None & None & Gapped \\
				\hline
				v & $\langle\cos(2\sqrt{2}\Theta_+)\rangle=1$ & $\langle\cos(2\sqrt{2}\Theta_-)\rangle=1$ & $\langle n_+^Y\rangle$ & None & Gapped \\
				\hline
				vi & None & $\langle\cos(\sqrt{2}\Phi_-)\rangle=-1$ & $\langle m_+^Y\rangle$ & $\langle n_-^X\rangle$, $\langle n_-^Z\rangle$ & Gapless \\
				\hline
				vii & None & $\langle\cos(2\sqrt{2}\Theta_-)\rangle=1$ & $\langle m_+^Y\rangle$ & $\langle n_+^Y\rangle$ & Gapless \\
				\bottomrule
			\end{tabular}
			\caption{Summary of phases in spin-ladder model under a longitudinal magnetic field ($B_y$). The ground state constraints of the symmetric and antisymmetric sectors are listed. We provide the leading and subleading order parameters for each phases as well as whether the excitation is gapped.}
			\label{tab:phases}
		\end{table*}

		In the presence of a longitudinal magnetic field, the two-leg spin ladder is described by $\hat{H}_++\hat{H}_-$, corresponding to two decoupled generalized sine-Gordon models. The ground states depend on the competition between $U_\pm$ and $V_\pm$, which induce a charge-density-wave-like (pinned $\Theta_\pm$) and superconductor-like (pinned $\Phi_\pm$) orders, respectively. In addition, the symmetric sector Hamiltonian, $\hat{H}_+$, also couples to $\zeta$ (related to the longitudinal magnetic field) and $\eta$ (parameter related to DM interactions in the leg). The complete phase diagram depends on $U_\pm$, $V_\pm$, $\zeta$, $\eta$, and $K_\pm$. Note that the analysis here is in the vicinity of zero magnetic field, where the Luttinger parameters are essentially the same as the isotropic AF Heisenberg chain, i.e., $K_\pm\approx1/2$ \cite{GiamarchiT2003}. For a sufficiently large magnetic field, one should expect a fully polarized state along the field direction, regardless of other parameters.
		
		A particularly interesting limit is the Luther-Emery point ($K_+=K_-=1/2$) in which the bosonized Hamiltonian $\hat{H}_{\pm}$ can be mapped to an exactly solvable quadratic fermion theory \cite{SheltonDG1996,GiamarchiT2003}. Notably, the system here is close to $K_+=K_-=1/2$ provided that the rung coupling is much weaker than the couplings within each chain. A detailed discussion on the mapping can be found in Appendix.~\ref{App:LE}. Here, we summarize the main results. In the fermionized description, the $U_\pm$ term becomes to a single-particle backscattering term (e.g., $\psi_L^{\dagger}\psi_R$), and the $V_\pm$ term becomes to a pairing term (e.g., $\psi_L\psi_R$). The $U_\pm$ term tends to develop a single-particle band gap, corresponding to pinning of $\Theta_\pm$; the $V_\pm$ term tends to develop a pairing gap, corresponding to pinning of $\Phi_\pm$. Thus, the $U_\pm$ and $V_\pm$ generically compete with each other, as expected from the bosonized description. The longitudinal magnetic field (associated with $\zeta$) is fermionized as a chemical potential term, which suppresses only the gap due to the $U_+$ term. On the contrary, the DM interaction along the leg (associated with $\eta$) is fermionized as a current term, which suppresses only the gap due to the $V_+$ term. The transition lines can be computed exactly by solving the quadratic fermion theory numerically. Moreover, the theory can be further expressed in terms of Majorana fermions, allowing for clear physical insights of competition between $U_\pm$ and $V_\pm$ \cite{SheltonDG1996}. In the following, we construct and discuss the phase diagrams, using the Luther-Emery fermionization and assuming $K_+=K_-=1/2$.

		First, we discuss the special case with $\eta=0$ and $\zeta=0$. In such a situation, the phase diagram can be derived by comparing $U_+=\frac{J_\perp-\Gamma_\perp}{\pi}$ and $V_+=\Gamma_\perp$ for the symmetric sector and $U_-=\frac{J_\perp-\Gamma_\perp}{\pi}$ and $V_-=J_\perp$ for the antisymmetric sector. The sign of $\frac{J_\perp-\Gamma_\perp}{\pi}$ is important for the pinning condition of the bosonic field $\Theta_\pm$. In the symmetric sector, the ground state tends to minimize $\frac{J_\perp-\Gamma_\perp}{\pi}\cos(2\sqrt{2}\Theta_+)$ for $\Gamma_\perp/J_\perp<1/(\pi+1)$ and $\Gamma_\perp\cos(\sqrt{2}\Phi_+)$ for $\Gamma_\perp/J_\perp>1/(\pi+1)$. Similarly, in the antisymmetric sector, the ground state tends to minimize $J_-\cos(\sqrt{2}\Phi_-)$ for $\Gamma_\perp/J_\perp<\pi+1$ and $-\frac{J_\perp-\Gamma_\perp}{\pi}\cos(2\sqrt{2}\Theta_-)$ for $\Gamma_\perp/J_\perp>\pi+1$. With the ground state constraints above, we conclude that there are three phases tuned by $\Gamma_\perp/J_\perp$. For $\Gamma_\perp/J_\perp<1/(\pi+1)$, a nonmagnetic gapped phase (phase i) is realized. For $1/(\pi+1)<\Gamma_\perp/J_\perp<\pi+1$, a magnetic gapped phase carrying finite $\langle n_-^X\rangle$ (phase ii) is realized. For $\Gamma_\perp/J_\perp>\pi+1$, another nonmagnetic gapped phase (phase iii) is realized. The system becomes gapless at the transition points because $\Phi_\pm$ and $\Theta_\pm$ cannot simultaneously acquire finite expectation values. Phases i and iii are different due to the distinct ground state constraints. Based on the strong rung coupling analysis in Sec.~\ref{Sec:strong_rung}, phases i and iii are likely adiabatically connected to product states of Eq.~(\ref{Eq:psi_generic_Bx}a) with $\alpha=0$ ($\Gamma_\perp/J_\perp\rightarrow 0$) and $\alpha=\pi/2$ ($\Gamma_\perp/J_\perp\rightarrow \infty$), respectively. The properties of the phases are summarized in Table~\ref{tab:phases}.

		For $\zeta\neq 0$ and $\eta=0$, phase i becomes less stable and turns into phase ii for a sufficiently large $\zeta$, as shown in Fig.~\ref{Fig:Ladder_PD_zero}(a). Technically, the pinning of $\Theta_+$ is weakened by a finite $\zeta$, making room for the competing order (i.e., the $V_+$ term). For $\zeta=0$ and $\eta\neq 0$, the phases ii and iii become unstable and undergo transitions for a sufficiently large $\eta$ as shown in Fig.~\ref{Fig:Ladder_PD_zero}(b). Phase ii can be driven into phase i ($\Gamma_\perp/J_\perp<1$) or phase iv ($\Gamma_\perp/J_\perp>1$) depending on $\Gamma_\perp/J_\perp$. Phase iv is another nonmagnetic gapped phase, whose pinning condition of $\Theta_+$ is different from phase i Meanwhile, phase iii can be driven into phase v, a magnetic gapped phase with finite $\langle n_+^Y\rangle$. See Table~\ref{tab:phases} for detailed descriptions of phases iv and v. 
		
		For $\zeta\neq 0$ and $\eta\neq 0$, commensurate-incommensurate transitions \cite{PokrovskyVL1979} can happen, corresponding to disordering the orders in the symmetric sector. Note that the antisymmetric sector is completely unaffected by $\zeta$ and $\eta$. In Fig.~\ref{Fig:Ladder_PD_By}, we show three phase diagrams with three representative values of $\eta$. Two new phases, vi and vii, emerge. Within phases vi and vii, both cosine terms in $\hat{H}_+$ [given by Eq.~(\ref{Eq:H_+-}a)] are irrelevant due to large $\zeta$ and $\eta$, realizing Luttinger liquids in the symmetric sector. In addition to finite magnetization along longitudinal ($y$) direction, phase vi contains power-law correlation in $n_-^X$ and $n_-^Z$, while phase vii contains power-law correlation in $n_+^Y$. These power-law correlations are expected to be stabilized in finite systems. The phase boundaries in Fig.~\ref{Fig:Ladder_PD_By} are computed based on the Luther-Emery quadratic fermion theory, as discussed previously and in Appendix~\ref{App:LE}. Note that the boundary between 
		phases i and ii in Fig.~\ref{Fig:Ladder_PD_By}a is not a vertical line. As a result, phase i can be driven into phase ii by the longitudinal field ($\zeta$). The same is true for the boundary between phases iv and ii [Fig.~\ref{Fig:Ladder_PD_By}b] and the boundary between phases v and iii [Fig.~\ref{Fig:Ladder_PD_By}c].

		Finally, we note that phases i, ii, iii, iv, and v do not have finite longitudinal magnetization even under a finite longitudinal magnetic field. The nonmagnetic phases i, iv, and v are protected by the gaps due to pinning of $\Theta_+$, which can be suppressed by the longitudinal magnetic field, corresponding to the commensurate-incommensurate transition \cite{PokrovskyVL1979}. Phases ii and iii are robust against the applied longitudinal field because pinning of $\Phi_+$ results in completely disordered $\Theta_+$, a consequence of the commutation relation between $\Theta_+$ and $\Phi_+$. Nevertheless, we expect a full spin polarization along $y$ for a sufficiently large longitudinal field regardless of $\Gamma_\perp/J_\perp$, which is not captured by our analysis in the vicinity of zero magnetic field.

		\subsection{Phase diagram with a transverse magnetic field}
		
		Now, we discuss the two-leg spin ladder phase diagram under a transverse magnetic field (i.e., $\vec{B}\perp\hat{y}$). In the weak rung coupling limit, we can construct the phase diagram based on the single chain model under a transverse field. Unlike the longitudinal case where Luttinger liquid generically manifests, the transverse field can induce gapped states \cite{GangadharaiahS2008,GarateI2010}. In the following, we construct low-energy phases of the two-leg spin ladder by using the single-chain results and treating the rung coupling as perturbations. Note that all the coordinates discussed in this subsection are related to the physical orientation as shown in Fig.~\ref{Fig:SL}, not the rotated basis. 
		
		The single-chain model realizes an isotropic AF Heisenberg model after gauging away the DM interaction along the chain.
		In this case, an arbitrarily small transverse field can induce a staggered magnetization along the longitudinal direction ($y$) in addition to a finite uniform magnetization in the field direction \cite{GarateI2010}. For $\vec{B}\parallel\hat{x}$, we construct a mean-field spin configurations as follows:
		\begin{align}\label{Eq:spin_conf_Bx_decoupled}
			\vec{s}_n=m_0\hat{x}+n_1(-1)^{n}\hat{y},\,\,\,\vec{\tau}_n=m_0\hat{x}+n_2(-1)^{n}\hat{y},
		\end{align}
		where $m_0$ denotes the uniform magnetization and $n_1,n_2$ denote the staggered magnetizations with $|n_1|=|n_2|$. To understand the phase diagram of the spin ladder under a $B_x$ field, we perform first-order perturbation theory in the rung coupling Hamiltonian $\hat{H}_{\text{rung}}$ [Eq.~(\ref{Eq:H_spin_ladder}b)], taking the mean-field spin configurations in Eq.~(\ref{Eq:spin_conf_Bx_decoupled}) as unperturbed wavefunctions. The energy correction due to the rung Hamiltonian $\hat{H}_{\text{rung}}$ is given by
		\begin{align}
			\delta E=\sum_n\left[(J_\perp+\Gamma_\perp)m_0^2+(J_\perp-\Gamma_\perp)n_1n_2\right].
		\end{align}
		$\delta E$ is minimized when $(J_\perp-\Gamma_\perp)n_1n_2<0$. Thus,
		we conclude that $\langle m_+^X\rangle\neq0$ and $\langle n_-^Y\rangle\neq0$ for $J_\perp>\Gamma_\perp$, and $\langle m_+^X\rangle\neq0$ and $\langle n_+^Y\rangle\neq0$ for $J_\perp<\Gamma_\perp$. Finally, for a sufficiently large $B_x$ field, the spin ladder model is fully polarized along the $x$ direction. 
		
		The $B_z$ transverse field case gives the same phases with $x$ and $z$ swapped. The coincidence of the two cases is not obvious as the rung coupling Hamiltonian $\hat{H}_{\text{rung}}$ [Eq.~(\ref{Eq:H_spin_ladder}b)] is anisotropic. Within our analysis, we find that the phases are determined by the $(J_\perp-\Gamma_\perp)s_n^y\tau_n^y$ term in $\hat{H}_{\text{rung}}$ [Eq.~(\ref{Eq:H_spin_ladder}b)], the same as the $B_x$ field case. Thus, we predict $\langle m_+^Z\rangle\neq0$ and $\langle n_-^Y\rangle\neq0$ for $J_\perp>\Gamma_\perp$, and $\langle m_+^Z\rangle\neq0$ and $\langle n_+^Y\rangle\neq0$ for $J_\perp<\Gamma_\perp$.

		\subsection{Finite-size systems}\label{Sec:Finite-size}
		
		The analysis in the weak rung coupling limit is based on continuum theory and bosonization, assuming an infinitely long ladder. However, the current Ge hole spin qubit experiments have only realized small systems such as $2\times 4$ quantum dot arrays \cite{ZhangX2025,HsiaoTK2024,FarinaPC2025a,JirovecD2025}. Therefore, it is important to examine whether predictions based on infinite systems survive the finite-size effects. In the following, we discuss complications in finite systems and focus mainly on the phases with a longitudinal magnetic field (along the $y$ direction).
		
		To realize an order predicted in the thermodynamic limit, the system size must be larger than the magnetic unit cell and the correlation length of the associated order. The former condition is easy, as the phases we predicted have the largest unit cell of two rungs. We then focus on the correlation length in each phase. The gapped phases (phases i-v in Table~\ref{tab:phases}) are generally more stable against finite-size effects. For small systems, one can, in principle, tune up the rung couplings (resulting in a large gap) such that the condition is fulfilled. Particularly, phases i and iii can be viewed as product states of (generalized) rung singlets, which should be realized in small quantum dot arrays such as the $2\times 4$ systems. 
		
		The gapless phases (phases vi and vii in Table~\ref{tab:phases}) are harder to confirm in finite-size systems. First, the finite length always induces finite-size gaps. Second, the order parameters $n^Y_+$, $n^X_-$, and $n^Z_-$ are fluctuating and power-law correlated in the infinite-size limit. These orders are stabilized in finite-size systems with sufficiently long lengths, but we generally expect a much larger length for the gapless phases compared to the gapped phases. 
		
		Another issue is the length scales associated with $\zeta$ (the longitudinal magnetic-field-induced spatial oscillation) and $\eta$ (the SOC-induced spatial oscillation). Several orders are predicted based on driving the system into incommensurate phases. Specifically, phase vi is driven by $\zeta$, phases iv and v are driven by $\eta$, and phase vii requires both $\zeta$ and $\eta$. To drive into the incommensurate phases, the corresponding length scale must be smaller than the system length. Increasing the longitudinal magnetic field can enhance $\zeta$ and, in principle, fulfill the condition $\pi/(2\zeta)<L$ with $L$ being the system length. Meanwhile, the value of $\eta$ is less tunable, and realizing phases iv, v, and vii requires $a\pi/\eta<L$. 
		
		Finally, realistic systems are typically under open boundary conditions, rather than the periodic boundary conditions often used in analytical approaches. The boundary condition determines the allowed discrete momentum points in a finite system. Moreover, the boundary spins experience half of the interactions, with different energetic conditions than the ``bulk'' spins. The predictions based on the infinite-size limit are strictly valid when the number of rungs is much larger than two.

		\section{Discussion}\label{Sec:Discussion}
	
		In this work, motivated primarily by recent experimental developments in fabricating Ge hole quantum dot based spin qubits with strong SOC as well as inter-dot exchange coupling, we study a two-leg spin ladder made of spin-orbit-coupled quantum dots, using two complementary approaches in the strong and weak rung coupling limits. In the strong rung coupling limit, we systematically derive the effective spin-$1/2$ model and show how to engineer the effective Hamiltonian by tuning the microscopic couplings as well as the magnetic field. In the weak rung coupling limit with identical legs, we employ the Abelian bosonization and Luther-Emery fermionization \cite{GogolinAO2004,GiamarchiT2003} and construct low-energy phase diagrams in several analytically tractable limits. Our work bridges the quantum many-body theory and the spin qubit device physics, and demonstrates the spin-orbit-coupled quantum dot spin ladders as an ideal platform for emulating exotic spin models, controlling quantum many-body states, and paving the way for quantum simulations and computations.

		The main distinction between our work and the conventional AF Heisenberg spin ladders \cite{GogolinAO2004,GiamarchiT2003} is the SOC-induced interactions, i.e., the DM and anisotropic exchange interactions. One notable technical advance here is the incorporation of SOC-induced interactions in two complementary limits. In the strong rung coupling limit, these SOC-induced interactions along the legs are projected onto the reduced Hilbert space, resulting in a rich set of inter-rung interactions that are absent in the conventional AF Heisenberg spin ladders \cite{GogolinAO2004,GiamarchiT2003}. In the weak rung coupling limit, the SOC-induced interactions along the leg can be gauged away by performing local basis rotations, and the rotated angle $\eta$ encodes the strength of SOC. Several new phases are generated due to the Anisotropic exchange interactions on the rungs and the commensurate-incommensurate transition driven by the longitudinal magnetic field ($y$ direction) as well as the SOC along the legs (encoded by $\eta$).

		The results in the strong rung coupling limit are closely related to the singlet-triplet encoding, which has been studied experimentally in the $2\times 4$ Ge hole quantum dot arrays \cite{ZhangX2025,JirovecD2025,FarinaPC2025a}. The strong rung coupling approach does not require the thermodynamic limit or continuum approximation, making it suitable for already-existing finite-size experimental systems. Different effective spin chain Hamiltonians can be realized by changing merely the orientation of the applied magnetic field. In addition, the DM interactions in the effective Hamiltonian can be fully suppressed in several cases, while the microscopic DM interactions are generically inevitable. Remarkably, a staggered DM interaction, which allows for a field-induced gap pertinent to the sine-Gordon model \cite{OshikawaM1997,OrignacE2007}, can be realized in the effective spin model through manipulating the microscopic spin-spin couplings, as we discussed in Sec.~\ref{Sec:strong_rung}. Our results suggest that Ge hole quantum arrays are an effective platform for simulating quantum spin models with DM interactions, an unprecedented direction in the artificially constructed quantum simulators.
		
		In the weak rung coupling limit, we construct the low-energy phase diagrams and reveal the subtle role of the DM interactions. These phase diagrams can provide hints for identifying the strength of SOC in the experiments. Specifically, the phase diagrams in Fig.~\ref{Fig:Ladder_PD_By} show strong dependence on $\eta$, the rotated phase variable due to DM interactions in the legs. Probing the many-body phase diagram of the spin ladder made of Ge hole spin qubits may constrain the possible range of the SOC-induced interactions, independent of how spin-flip tunnelings arise.  
		
		One major approximation in the weak coupling analysis is the identical-leg assumption. Relaxing this condition, the differences in the rotated phases, $\eta_s-\eta_\tau$, and the asymmetry in $g$ factors generically contribute to the zero modes of the $\Phi_-$ and $\Theta_-$ variables, respectively. As a result, commensurate-incommensurate transitions \cite{PokrovskyVL1979} can also happen in the chain-antisymmetric sectors, realizing more complicated phase diagrams. A systematic study for the generic non-identical legs cases will be an interesting future direction. We also note that in the extreme situation that $J'=D'=\Gamma'=0$ (the Kondo-necklace-like limit), the bosonization approach becomes inadequate because one of the legs cannot be described by a Luttinger liquid. Otherwise, the generic non-identical legs cases can be studied by extending the theoretical framework here, provided that both legs can be described by the Luttinger liquid type of theory.

		Now, we compare the results of the two complementary approaches in this work. Such comparisons are not obvious because the strong rung coupling approach primarily constructs an effective Hamiltonian (without analyzing the phases directly), and the weak rung coupling approach predicts low-energy phases in the thermodynamic limit utilizing bosonization. In addition, the adiabaticity between two limits is not always guaranteed. Regardless of the issues mentioned above, we point out that phases i and iii are likely adiabatically connected to the product states of $\psi_1$ given by Eq.~(\ref{Eq:psi_generic_Bx}a) with $\alpha=0$ and $\alpha=\pi/2$, respectively. To support this speculation, we compare the expectation values of $s^\mu\tau^\mu$ (in the physical basis, not the rotated basis) and find qualitative consistency between the two distinct approaches. The results also suggest that the sizes of phases i and iii shrink as the rung couplings are enhanced, and two phases are reduced to two points in the strong rung coupling limits. Thus, we speculate that there are multiple phase transitions driven by the rung couplings, which should be examined further experimentally or in numerical simulations (e.g., DMRG).

		We conclude by discussing several experimentally relevant issues. 
		Unlike many other spin qubit platforms \cite{BurkardG2023}, Ge contains only a small percentage of isotopes that possess nonzero nuclear spins. Thus, the magnetic noise is not the most significant source of decoherence \cite{ScappucciG2021}. Meanwhile, the charge noise, strong SOC, and $g$ factors are expected to be the considerable origins of the decoherence \cite{ScappucciG2021}. These effects limit the time scale during which the systems remain coherent. Next, we discuss the disorders in the realistic experiments.
		The Ge hole planar system has strong anisotropy in the $g$ factor \cite{ScappucciG2021}, and the primary source of disorder is the variation of the $g$ factor, which has been neglected completely in this work. It is important to investigate how the $g$-factor variation, which primarily acts as the random-field disorder in the spin models, affects the phase diagram constructed in this work. Moreover, the $g$-factor variation also can result in the Zeeman energy difference between $s$ and $\tau$ spins, thereby modifying the eigenstates of $\hat{H}_n$ [Eq.~(\ref{Eq:H_n})]. Besides the $g$-factor variation, the disorder can also impact the onsite energy inside a quantum dot and the inter-dot tunnelings, resulting in inhomogeneity in spin-spin interactions. In principle, such spatial fluctuations can be minimized by fine tuning the plunger and barrier gates in the experiments. These corrections, while assumed to be small in this work, deserve more detailed investigations in future studies. Additionally, distinguishing different quantum phases in a finite-size system can be subtle and challenging, as we discussed in Sec.~\ref{Sec:Finite-size}. Therefore, developing experimentally accessible diagnostics, e.g., analyzing low-energy excitations, may provide signatures of the underlying phases. Such protocols will be invaluable for connecting theoretical predictions to future Ge hole quantum dot spin ladder experiments. We anticipate that the theoretical frameworks and results presented in this work will serve as a foundation for designing, controlling, and characterizing forthcoming Ge hole quantum dot ladder experiments.

		\begin{acknowledgments}
			We thank Eugene Demler, Silas Hoffman, Katharina Laubscher, Masaki Oshikawa, Jay D. Sau, and Zhentao Wang for useful discussion. This work is supported by the Laboratory for Physical Sciences.
		\end{acknowledgments}
	
	\appendix

	\section{Staggered DM interaction in effective spin chains}\label{App:Staggered_DM}
	
	In this Appendix, we show how to engineer a staggered DM interaction. Specifically, we consider the spin ladder model with a magnetic field along $x$ direction with the low-energy wavefunctions given by $\psi_1=\frac{1}{\sqrt{2}}\left(\ket{\uparrow_s\downarrow_{\tau}}-\ket{\downarrow_s\uparrow_{\tau}}\right)$ and $\psi_2=\frac{1}{2}\left(\ket{\uparrow_s\uparrow_{\tau}}+\ket{\uparrow_s\downarrow_{\tau}}+\ket{\downarrow_s\uparrow_{\tau}}+\ket{\downarrow_s\downarrow_{\tau}}\right)$. With the low-energy states and Eq.~(\ref{Eq:Local_op}), we project the one-spin local operators onto the reduce Hilbert space as follows:
	\begin{subequations}
		\begin{align}
			s^x_n\rightarrow\frac{1}{4}\left(I_n-\sigma_n^z\right),\,\,\,s^y_n\rightarrow\frac{1}{2\sqrt{2}}\sigma_n^y,\,\,\,s^z_n\rightarrow\frac{1}{2\sqrt{2}}\sigma_n^x,\\
			\tau^x_n\rightarrow\frac{1}{4}\left(I_n-\sigma_n^z\right),\,\,\,\tau^y_n\rightarrow\frac{-1}{2\sqrt{2}}\sigma_n^y,\,\,\,\tau^z_n\rightarrow\frac{-1}{2\sqrt{2}}\sigma_n^x.
		\end{align}
	\end{subequations}

	Now, we consider a leg Hamiltonian [which is different from Eq.~(\ref{Eq:H_spin_ladder}a)] as follows:
	\begin{align}
		\nonumber\hat{H}_{\text{leg}}'=&\sum_n\left[\begin{array}{c}
			J_n\vec{s}_{n}\cdot\vec{s}_{n+1}+D_n\hat{y}\cdot\left(\vec{s}_{n}\times\vec{s}_{n+1}\right)\\[1mm]
			+\Gamma_n\left(2 s^y_{n}s^y_{n+1}-\vec{s}_{n}\cdot\vec{s}_{n+1}\right)
		\end{array}\right]\\
		&+\sum_n\left[\begin{array}{c}
			J'_n\vec{\tau}_{n}\cdot\vec{\tau}_{n+1}+D'_n\hat{y}\cdot\left(\vec{\tau}_{n}\times\vec{\tau}_{n+1}\right)\\[1mm]
			+\Gamma'_n\left(2 \tau^y_{n}\tau^y_{n+1}-\vec{\tau}_{n}\cdot\vec{\tau}_{n+1}\right)
		\end{array}\!\right],
	\end{align}
	where 
	\begin{align}
		(J_n,D_n,\Gamma_n)=&\begin{cases}
			(J,D,\Gamma)\text{ for even }n,\\
			(0,0,0)\text{ for odd }n,
		\end{cases}\\
		(J_n',D_n',\Gamma_n')=&\begin{cases}
			(0,0,0)\text{ for even }n,\\
			(J,D,\Gamma)\text{ for odd }n,
		\end{cases}
	\end{align}
	
	Using the above expressions, we derive the contribution from $\hat{H}_{\text{leg}}'$:
	\begin{align}\label{Eq:H_leg'}
		\hat{H}_{\text{leg}}'\rightarrow\sum_n\left\{\begin{array}{c}
			\!\!\frac{(J-\Gamma)}{8}\left[\frac{(I_n-\sigma_n^z)(I_{n+1}-\sigma_{n+1}^z)}{2}+\sigma_n^x\sigma_{n+1}^x\right]\\[2mm]
			+\frac{(J+\Gamma)}{8}\sigma_n^y\sigma_{n+1}^y\\[2mm]
			\!\!+\frac{(-1)^nD}{8\sqrt{2}}\left[\sigma_n^x(I_{n+1}-\sigma^z_{n+1})-(I_n-\sigma^z_n)\sigma_{n+1}^x\right]
		\end{array}
		\right\}.
	\end{align}
	The last line of the above curly can be organized into (ignoring the boundary terms for now)
	\begin{align}
		\frac{D}{8\sqrt{2}}\sum_n(-1)^n\left[\left(\sigma_n^z\sigma_{n+1}^x-\sigma_n^x\sigma_{n+1}^z\right)+2\sigma_n^x
		\right].
	\end{align}
	The staggered transverse field term comes from the one-spin local operator in $s_n^x$ and $\tau_n^x$, which do not contribute to the effective model at all for $\langle\sigma_n^z\rangle=1$, corresponding to the singlet ground state. As a result, the staggered DM and the staggered transverse field interactions cancel each other in the $\langle\sigma_n^z\rangle=1$ sector.
	
	With the results above, we can map the two-leg spin ladder into an effective spin-$1/2$ model (with $\mathcal{S}_n^{\mu}=\sigma_n^{\mu}/2$) given by
	\begin{align}\label{Eq:H_eff_St_DM}
		\nonumber\hat{H}_{\text{eff}}=&\sum_n\left[\begin{array}{c}
			\tilde{J}_x\mathcal{S}_n^x\mathcal{S}_{n+1}^x+\tilde{J}_y\mathcal{S}_n^y\mathcal{S}_{n+1}^y+\tilde{J}_z\mathcal{S}_n^z\mathcal{S}_{n+1}^z\\[1mm]
			+\tilde{D}_y(-1)^n\left(\mathcal{S}_n^z\mathcal{S}_{n+1}^x-\mathcal{S}_n^x\mathcal{S}_{n+1}^z\right)\\[1mm]
			+\tilde{D}_y(-1)^n\mathcal{S}_n^x-\tilde{h}_z\mathcal{S}^z_n
		\end{array}
		\right]\\
		&+\frac{\tilde{J}_z}{2}\left(\mathcal{S}_{n_L}+\mathcal{S}_{n_R}\right)+\frac{\tilde{D}_y}{2}\left[(-1)^{n_L}\mathcal{S}^x_{n_L}-(-1)^{n_R}\mathcal{S}^x_{n_R}\right],
	\end{align}
	where $n_R$ and $n_L$ are the right and left boundary sites,
	\begin{subequations}
		\begin{align}
			\tilde{J}_x=&(J-\Gamma)/2,\\
			\tilde{J}_y=&(J+\Gamma)/2,\\
			\tilde{J}_z=&(J-\Gamma)/4,\\
			\tilde{D}_y=&D/2\sqrt{2},\\
			\tilde{h}_z=&J_\perp-g_xB_x+(J-\Gamma)/4.
		\end{align}
	\end{subequations}
	
	Equation~(\ref{Eq:H_eff_St_DM}) is slightly different from the original proposals of the staggered DM interaction \cite{OrignacE1997,AffleckI1999} due to the existence of the staggered transverse field term. Such a staggered field tends to cancel the staggered DM contribution for $\tilde{h}_z>0$ but enhances the staggered DM contribution for $\tilde{h}_z<0$. Technically, this spin-dependent contribution originates from the local one-spin operator projection, $s^x_n,\tau_n^x\rightarrow\left(I_n-\sigma_n^z\right)/4$. Specifically, the last line in the curly brackets of Eq.~(\ref{Eq:H_leg'}) explicitly shows the absence of contributions for the up-spin state in the $\sigma$ spins.

	\begin{widetext}

	\section{Bosonized expressions of spin-spin interactions on the rungs}\label{App:Bosonized_interaction}

		In this appendix, we discuss the derivations of Eq.~(\ref{Eq:H_bosonized_12_original}). First, the spin-spin interactions along the legs can be bosonized into a quantum sine-Gordon model as discussed in Refs.~\cite{GogolinAO2004,GiamarchiT2003}. Here, we focus on the spin-spin interactions on the rungs. Using Eq.~(\ref{Eq:Bosonization_s_tau}), the spin-spin interactions are expressed by
		\begin{align}
			\nonumber s^+(x)\tau^-(x)=&\left\{\frac{1}{\sqrt{2\pi a}}e^{i\phi_1(x)}\left\{(-1)^{x/a}+\cos\left[2\theta_1(x)\right]\right\}\right\}\left\{\frac{1}{\sqrt{2\pi a}}e^{-i\phi_2(x)}\left\{(-1)^{x/a}+\cos\left[2\theta_2(x)\right]\right\}\right\}\\
			=&\frac{e^{i\left[\phi_1(x)-\phi_2(x)\right]}}{2\pi a}\left\{1+\cos\left[2\theta_1(x)\right]\cos\left[2\theta_2(x)\right]+(-1)^{x/a}\Big\{\cos\left[2\theta_1(x)\right]+\cos\left[2\theta_2(x)\right]\Big\}
			\right\}\\
			\nonumber s^+(x)\tau^+(x)=&\left\{\frac{1}{\sqrt{2\pi a}}e^{i\phi_1(x)}\left\{(-1)^{x/a}+\cos\left[2\theta_1(x)\right]\right\}\right\}\left\{\frac{1}{\sqrt{2\pi a}}e^{i\phi_2(x)}\left\{(-1)^{x/a}+\cos\left[2\theta_2(x)\right]\right\}\right\}\\
			=&\frac{e^{i\left[\phi_1(x)+\phi_2(x)\right]}}{2\pi a}\left\{1+\cos\left[2\theta_1(x)\right]\cos\left[2\theta_2(x)\right]+(-1)^{x/a}\Big\{\cos\left[2\theta_1(x)\right]+\cos\left[2\theta_2(x)\right]\Big\}
			\right\}\\
			\nonumber s^+(x)\tau^z(x)=&\left\{\frac{1}{\sqrt{2\pi a}}e^{i\phi_1(x)}\left\{(-1)^{x/a}+\cos\left[2\theta_1(x)\right]\right\}\right\}\left\{\frac{1}{\pi}\partial_x\theta_2(x)+\frac{1}{\pi a}(-1)^{x/a}\cos\left[2\theta_2(x)\right]\right\}\\
			\nonumber=&\frac{e^{i\phi_1(x)}}{\pi\sqrt{2\pi a}}\left[\partial_x\theta_2(x)\right]\cos\left[2\theta_1(x)\right]+\frac{e^{i\phi_1(x)}}{\pi a\sqrt{2\pi a}}\cos\left[2\theta_2(x)\right]\\
			&
			+\frac{(-1)^{x/a}e^{i\phi_1(x)}}{\pi\sqrt{2\pi a}}\Big\{\left[\partial_x\theta_2(x)\right]+\cos\left[2\theta_2(x)\right]\cos\left[2\theta_1(x)\right]\Big\},\\
			\nonumber s^z(x)\tau^+(x)=&\left\{\frac{1}{\pi}\partial_x\theta_1(x)+\frac{1}{\pi a}(-1)^{x/a}\cos\left[2\theta_1(x)\right]\right\}\left\{\frac{1}{\sqrt{2\pi a}}e^{i\phi_2(x)}\left\{(-1)^{x/a}+\cos\left[2\theta_2(x)\right]\right\}\right\}\\
			\nonumber=&\frac{e^{i\phi_2(x)}}{\pi\sqrt{2\pi a}}\left[\partial_x\theta_1(x)\right]\cos\left[2\theta_2(x)\right]+\frac{e^{i\phi_2(x)}}{\pi a\sqrt{2\pi a}}\cos\left[2\theta_1(x)\right]\\
			&
			+\frac{(-1)^{x/a}e^{i\phi_2(x)}}{\pi\sqrt{2\pi a}}\Big\{\left[\partial_x\theta_1(x)\right]+\cos\left[2\theta_1(x)\right]\cos\left[2\theta_2(x)\right]\Big\},\\
			\nonumber s^z(x)\tau^z(x)=&\left\{\frac{1}{\pi}\partial_x\theta_1(x)+\frac{1}{\pi a}(-1)^{x/a}\cos\left[2\theta_1(x)\right]\right\}\left\{\frac{1}{\pi}\partial_x\theta_2(x)+\frac{1}{\pi a}(-1)^{x/a}\cos\left[2\theta_2(x)\right]\right\}\\
			\nonumber=&\frac{1}{\pi^2}\left[\partial_x\theta_1(x)\right]\left[\partial_x\theta_2(x)\right]+\frac{1}{\pi^2 a^2}\cos\left[2\theta_1(x)\right]\cos\left[2\theta_2(x)\right]\\
			&+\frac{(-1)^{x/a}}{\pi^2 a}\Big\{\left[\partial_x\theta_1(x)\right]\cos\left[2\theta_2(x)\right]+\left[\partial_x\theta_2(x)\right]\cos\left[2\theta_1(x)\right]\Big\}.
		\end{align}
		The expressions above contain multiple terms with different scaling dimensions. Moreover, there are some terms with a $(-1)^{x/a}$ factor, indicating the highly oscillatory nature. For translation invariant systems, we can ignore the terms with $(-1)^{x/a}$ as these terms are parametrically small due to the oscillation. We also neglect terms that are unlikely to be relevant in our analysis. For example, we use $s^+\tau^-\approx e^{i\left[\phi_1-\phi_2\right]}/(2\pi a)$ for deriving Eq.~(\ref{Eq:H_bosonized_12_original}), ignoring the $e^{i\left[\phi_1-\phi_2\right]}\cos(\theta_1)\cos(\theta_2)$ term and the terms with $(-1)^{x/a}$.

	\end{widetext}

	\section{Luther-Emery fermionization and phase diagram construction}\label{App:LE}

	Here, we discuss the Luther-Emery limit, $K_+=K_-=1/2$, allowing for exact analysis of the spin ladder problem. First, we define $\bar{\Theta}_{\pm}=\sqrt{2}\Theta_{\pm}$ and $\bar{\Theta}_{\pm}=\Theta_{\pm}/\sqrt{2}$. The Hamiltonian $\hat{H}_{\pm}$ in Eq.~(\ref{Eq:H_+-}) becomes
	\begin{align}
		\nonumber\hat{H}_{\pm}\rightarrow&\int dx\frac{v_{\pm}}{2\pi}\left[\left(\partial_x\bar{\Phi}_{\pm}\right)^2+\left(\partial_x\bar{\Theta}_{\pm}\right)^2\right]\\
		\nonumber&+\frac{U_{\pm}}{2\pi a}\int dx\cos\left(2\bar{\Theta}_{\pm}+4\zeta_{\pm} x\right)\\
		&+\frac{V_{\pm}}{2\pi a}\int dx\cos\left(2\bar{\Phi}_{\pm}-2\eta_{\pm} x/a\right),
	\end{align}
	where $U_+=U_-=(J_\perp-\Gamma_{\perp})/\pi$, $V_+=\Gamma_\perp$, $V_-=J_\perp$, $\zeta_+=\zeta$, $\zeta_-=0$, $\eta_+=\eta$, and $\eta_-=0$. For spin ladders with asymmetric legs, $\zeta_-$ and $\eta_-$ can become nonzero, resulting in a more complicated situation than our identical leg case. The modulating factors can be expressed as derivatives of bosonic fields. Specifically, we shift the bosonic fields $\bar{\Theta}_\pm\rightarrow \bar{\Theta}_\pm-2\zeta_\pm x$ and $\bar{\Phi}_\pm\rightarrow \bar{\Phi}_\pm+\eta_{\pm} x/a$. The Hamiltonian $\hat{H}_{\pm}$ becomes
	\begin{align}
		\nonumber\hat{H}_{\pm}\rightarrow&\int dx\frac{v_{\pm}}{2\pi}\left[\left(\partial_x\bar{\Phi}_{\pm}\right)^2+\left(\partial_x\bar{\Theta}_{\pm}\right)^2\right]\\
		\nonumber&+\int dx\left[-\frac{2v\zeta_\pm}{\pi}\partial_x\bar{\Theta}_\pm+\frac{v\eta_\pm}{\pi a}\partial_x\bar{\Phi}_\pm\right]\\
		&+\int dx\left[\frac{U_{\pm}}{2\pi a}\cos\left(2\bar{\Theta}_{\pm}\right)+\frac{V_{\pm}}{2\pi a}\cos\left(2\bar{\Phi}_{\pm}\right)\right].
	\end{align}
	The above Hamiltonian is related to a quadratic fermion theory. First, we define the Luther-Emery fermions  \cite{GiamarchiT2003,GogolinAO2004} as follows:
	\begin{align}
		R_\pm=\frac{e^{-i\pi/2}}{\sqrt{2\pi a}}e^{i\left(\bar{\Phi}_{\pm}+\bar{\Theta}_{\pm}\right)},\,\,\, L_\pm=\frac{1}{\sqrt{2\pi a}}e^{i\left(\bar{\Phi}_{\pm}-\bar{\Theta}_{\pm}\right)}.
	\end{align}
	In the above expression, we have introduced a phase factor $-\pi/2$ in $R_\pm$, which is convenient for the following analysis.
	With these fermion fields, the $\hat{H}_{\pm}$ can be expressed by
	\begin{align}
		\nonumber\hat{H}_{\pm}=&\int dx \,v_{\pm}\left[R^{\dagger}_{\pm}\left(-i\partial_xR_{\pm}\right)-L^{\dagger}_{\pm}\left(-i\partial_xL_{\pm}\right)\right]\\
		\nonumber&-2\zeta_\pm v_\pm\int dx\left(R^{\dagger}_\pm R_\pm+L^{\dagger}_\pm L_\pm\right)\\
		\nonumber&+\frac{\eta_\pm v_\pm}{a}\int dx\left(R^{\dagger}_\pm R_\pm-L^{\dagger}_\pm L_\pm\right)\\
		&+\int dx\left[U_\pm\left(iL^{\dagger}_\pm R_\pm+\text{H.c.}\right)+V_\pm\left(iL_\pm R_\pm+\text{H.c.}\right)\right].
	\end{align}
	The above Hamiltonian can be diagonalized exactly under Fourier transform and the Bogoliubov-de-Gennes (BdG) formulation. 
	
	Notably, the gap dominated by $U_\pm$ can be suppressed by a finite $\zeta_\pm$, but not the $\eta_{\pm}$. A finite $\eta_{\pm}$ results in a shift of the $k$ points in the band edge without modifying the spectral gap. Similarly, the gap dominated by $V_\pm$ is only influenced by $\eta_\pm$, not the $\zeta_\pm$. The precise critical values for the commensurate-incommensurate transitions may require diagonalizing a $4\times 4$ BdG matrix, which is standard and omitted here.

	Alternatively, we can express the Hamiltonian in terms of Majorana fermions. Following Ref.~\cite{SheltonDG1996}, we rewrite the Hamiltonian $\hat{H}_{\pm}$ in terms of Majorana fermions, which are defined as follows:
	\begin{subequations}
		\begin{align}
			\chi_{R,\pm}=&\frac{R_{\pm}+R^{\dagger}_{\pm}}{\sqrt{2}},\,\,\rho_{R,\pm}=\frac{R_{\pm}-R^{\dagger}_{\pm}}{\sqrt{2}i},\\
			\chi_{L,\pm}=&\frac{L_{\pm}+L^{\dagger}_{\pm}}{\sqrt{2}},\,\,\rho_{L,\pm}=\frac{L_{\pm}-L^{\dagger}_{\pm}}{\sqrt{2}i}.
		\end{align}
	\end{subequations}
	The Hamiltonians become
	\begin{subequations}\label{Eq:H_pm_Majorana}
		\begin{align}
			\nonumber\hat{H}_{+}=&\int dx \left[\begin{array}{c}
				-\frac{iv_+}{2}\left(\chi_{R,+}\partial_x\chi_{R,+}-\chi_{L,+}\partial_x\chi_{L,+}\right)\\[2mm]
				+i(U_++V_+)\chi_{L,+}\chi_{R,+}
			\end{array}\right]\\
			\nonumber&+\int dx \left[\begin{array}{c}
				-\frac{iv_+}{2}\left(\rho_{R,+}\partial_x\rho_{R,+}-\rho_{L,+}\partial_x\rho_{L,+}\right)\\[2mm]
				+i(U_+-V_+)\rho_{L,+}\rho_{R,+}
			\end{array}\right]\\
			&+\int dx\left[\begin{array}{c}
				-iv\left(2\zeta-\frac{\eta}{a}\right)\chi_{R,+}\rho_{R,+}\\[1mm]
				-iv\left(2\zeta+\frac{\eta}{a}\right)\chi_{L,+}\rho_{L,+}
			\end{array}\right],\\
			\nonumber\hat{H}_{-}=&\int dx \left[\begin{array}{c}
				-\frac{iv_-}{2}\left(\chi_{R,-}\partial_x\chi_{R,-}-\chi_{L,-}\partial_x\chi_{L,-}\right)\\[2mm]
				+i(U_-+V_-)\chi_{L,-}\chi_{R,-}
			\end{array}\right]\\
			&+\int dx \left[\begin{array}{c}
				-\frac{iv_-}{2}\left(\rho_{R,-}\partial_x\rho_{R,-}-\rho_{L,-}\partial_x\rho_{L,-}\right)\\[2mm]
				+i(U_--V_-)\rho_{L,-}\rho_{R,-}
			\end{array}\right],
		\end{align}
	\end{subequations}
	The first two lines of Eqs.~(\ref{Eq:H_pm_Majorana}a) and (\ref{Eq:H_pm_Majorana}b) describe two decoupled Majorana fermions with gaps $U_\pm+V_\pm$ for the $\chi$ Majorana fermions and $U_\pm-V_\pm$ for the $\rho$ Majorana fermions. The last line of Eq.~(\ref{Eq:H_pm_Majorana}a) describes couplings between $\chi$ and $\rho$ Majorana fermions. The competition between $U_\pm$ and $V_\pm$ is clear in the Majorana basis, and the spectrum gap closes when $U_\pm-V_\pm$ or $U_\pm+V_\pm$ vanishes. The $\zeta$ and $\eta$ terms induce coupling between $\chi_{r,+}$ and $\rho_{r,+}$ with $r=R,L$.
	
	We comment on two different representations of $\hat{H}_\pm$. In the chiral fermion basis ($R$ and $L$), the roles of $\zeta$ and $\eta$ are transparent as they induce commensurate-incommensurate transitions. In the Majorana fermion basis ($\chi$ and $\rho$), the competition between $U_\pm$ and $V_\pm$ is transparent, and the interaction-driven transition happens when $|U_\pm|=|V_\pm|$. These two different representations help us understand the results from two complementary perspectives. Note that the two representations are equivalent and thus give the same results.

	Finally, we discuss how to construct the phase diagrams in Figs.~\ref{Fig:Ladder_PD_zero} and \ref{Fig:Ladder_PD_By}. In the absence of $\zeta$ and $\eta$, we obtain phases i, ii, and iii with the phase boundary directly determined by $U_\pm$ and $V_\pm$, as discussed above. In Fig.~\ref{Fig:Ladder_PD_zero}(a), the curve separating phases i and ii with $\zeta>0$ is determined by the closing of the spectral gap by tuning $\zeta$; we obtain the curve (not the vertical lines) in Fig.~\ref{Fig:Ladder_PD_zero}(b) by monitoring the precise value of $\eta$ such that the spectral gap closes. Similarly, the phase boundary between two gapped phases can be obtained through the same procedure in Fig.~\ref{Fig:Ladder_PD_By}. The phase boundary between a gapped phase and a gapless phase is determined by the value of $\zeta$ such that the spectral gap just closes in Fig.~\ref{Fig:Ladder_PD_By}. The phase boundaries in Figs.~\ref{Fig:Ladder_PD_zero} and \ref{Fig:Ladder_PD_By} are specific to the Luther-Emery point (which is believed to be close to our situation). We expect the qualitative phase diagram to remain valid for systems slightly tuned away from the Luther-Emery point.

	%%\bibliography{Spin_qubit}


\end{document}